\documentclass[12pt]{article}

\usepackage{bbold}
\usepackage{amssymb}
\usepackage{slashed}
\usepackage{graphics}
\usepackage{epsfig}

\declareslashed{}{/}{.1}{0}{E}

\def\Nt{{\cal N}}
\def\Tt{{\cal T}}

\def\vphi{\varphi}

\def\ss{\scriptstyle}
\def\sss{\scriptscriptstyle}
\def\Real{\mathbb R}
\def\G{{\hspace{.3mm}\bf g\hspace{.3mm}}}
\def\GRAD{{\hspace{.3mm}\bf grad\hspace{.3mm}}}
\def\DIV{{\hspace{.3mm}\bf div\hspace{.3mm}}}
\def\TR{{\hspace{.3mm}\bf tr\hspace{.3mm}}}
\def\N{{\hspace{.3mm}\bf N\hspace{.3mm}}}
\def\GG{{\hspace{.3mm}\bf G\hspace{.3mm}}}
\def\GGRAD{{\hspace{.3mm}\bf Grad\hspace{.3mm}}}
\def\SGRAD{{\hspace{.3mm}\bf {\cal G}rad\hspace{.3mm}}}
\def\DDIV{{\hspace{.3mm}\bf Div\hspace{.3mm}}}
\def\SDIV{{\hspace{.3mm}\bf {\cal D}iv\hspace{.3mm}}}
\def\TTR{{\hspace{.3mm}\bf Tr\hspace{.3mm}}}
\def\N{{\hspace{.3mm}\bf N\hspace{.3mm}}}
\def\c{{\hspace{.3mm}\bf c\hspace{.3mm}}}

\def\GAMMA{{\hspace{.3mm}\bf {\gamma} \hspace{.3mm}}}
\def\GAMMAC{{\hspace{.3mm}\bf \gamma^{*} \hspace{.3mm}}}
\def\DSLASH{\slashed{D}}

\def\D{{\hspace{.3mm}\cal D\hspace{.3mm}}}

\def\D{{\cal D}}

\def\r{\rho}
\def\s{\sigma}

\def\wt#1{\widetilde{#1}}

\def\ol#1{\overline{#1}} 
\def\sss{\scriptscriptstyle}

\def\d{\partial}
\def\m{\mu}
\def\n{\nu}

\def\vphi{\varphi}

\def\be{\begin{equation}}
\def\ee{\end{equation}}
\def\beq{\begin{equation}}
\def\eeq{\end{equation}}
\def\bea{\begin{eqnarray}}
\def\eea{\end{eqnarray}} 
\def\beqa{\begin{equation}\begin{array}{l}}
\def\eeqa{\end{array}\end{equation}}

\def\eqn#1{(\ref{#1})}

\def\eqref#1{eq.~(\ref{eq:#1})}

  \def\g{\gamma}

\def\L{{\it\Lambda}}

\def\vphi{\varphi}

\def\nn{\nonumber}

\begin{document}

\thispagestyle{empty}

\vspace{.8cm}
\setcounter{footnote}{0}
\begin{center}
{\Large
 {\bf Constant Curvature Algebras and Higher Spin Action 
Generating Functions}\\[10mm]

 {\sc \small K.~Hallowell and A.~Waldron\\[6mm]}

 {\em\small Department of Mathematics, University of California,
            Davis CA 95616, USA\\ 
            {\tt hallowell,wally@math.ucdavis.edu}}\\[5mm]
}

\bigskip

\bigskip

{\sc Abstract}\\
\end{center}

{\small
\begin{quote}

The algebra of differential geometry operations on symmetric 
tensors over constant  curvature manifolds forms a novel 
deformation of the $sl(2,\Real)\, \raisebox{.6mm}{$\sss |$}
\hspace{-2.15mm}\times\Real^2$ Lie algebra.
We present a simple calculus for calculations in its universal
enveloping algebra. As an application, we derive generating
functions for the actions and gauge invariances of massive, partially massless
and massless (for both bose and fermi statistics)
higher spins on constant curvature backgrounds.
These are formulated in terms of a minimal set of covariant, unconstrained,
fields rather than towers of auxiliary fields. 
Partially massless gauge transformations are shown to arise as
degeneracies of the flat, massless gauge transformation in
one dimension higher.
Moreover, our results and calculus offer a considerable simplification
over existing techniques for handling higher spins. 
In particular, we show how theories of arbitrary spin
in dimension $d$
can be rewritten in terms of a single scalar field in
dimension $2d$ where the $d$ additional dimensions
correspond to coordinate differentials.
We also develop
an analogous framework for spinor-tensor fields in 
terms of the corresponding superalgebra.

\bigskip

\bigskip

\end{quote}
}

\newpage

\vfill

\tableofcontents

\vfill


\section{Introduction}

The central objects of our study are sections of the symmetric
tensor bundle $\odot T^*M$ for constant curvature manifolds $M$. 
In physics parlance, these are totally symmetric tensor
fields whose most common application is to theories of higher
spin particles, and our physical motivation is to provide
a formulation of these theories in terms of a simple calculus
of operators acting on symmetric tensors\footnote{The tensorial
side of higher spin computations is necessarily complicated. For
a discussion of some of the difficulties involved, see~\cite{Bandos:2005mb}.}. 
The simplicity of the
algebra we find suggests that totally symmetric tensors, 
just like their totally antisymmetic differential form counterparts, 
may be useful tools for mathematical studies of constant curvature
manifolds.

Our main results are rather simple, so we summarize them here in 
the Introduction, leaving detailed derivations, explanations and comments to 
the ensuing Sections of this Article:

A first and key step is to introduce the index operator $\N$ 
whose eigenvectors are sections $\Phi\in\odot T^*M$ with a definite
number of indices, ${\it i.e.}$
\be
\N\Phi=s\Phi\quad \Longrightarrow\quad \Phi=\vphi_{\mu_1\ldots\mu_s}\ 
dx^{\mu_1}\ldots dx^{\mu_s}\, ,
\ee
where $dx^{\mu_1}\ldots dx^{\mu_s}\equiv\frac{1}{s!}\sum_{\sigma}
dx^{\mu_{\sigma(1)}}\otimes\cdots\otimes dx^{\mu_{\sigma(s)}}$.
This is in keeping with our physics goal of writing generating
functions for theories of arbitrary spin~$s$. 

Next we introduce operators
$\G,\TR:\odot T^*M\rightarrow\odot T^*M$ which act on eigenvectors of $\N$
by multiplication by the metric and symmetrizing, and contraction of 
a pair of indices, respectively. The triplet $\{\G,\N,\TR\}$ generate
the Lie algebra $sl(2,\Real)$. The doublet formed from
the gradient and divergence operators
$\{\GRAD,\DIV\}$ then transform as the fundamental representation.
Were these operators to commute, we would be faced with the Lie algebra
 $sl(2,\Real)\, \raisebox{.6mm}{$\sss |$}
\hspace{-2.15mm}\times\Real^2$, however, instead we have a deformation
thereof because on a constant curvature manifold,
\be
[\DIV,\GRAD]=\square-2\c\, .
\ee
This relation completes our ``constant curvature algebra''.
The operator~$\square$ is central and related to the usual Laplacian 
$\Delta$ by
\be
\square=\Delta+\c=[\DIV,\GRAD]+2\c\, ,
\ee
and 
\be
\c=\G\TR-\N(\N+n-1)
\ee 
is the $sl(2,\Real)$ Casimir while $n\equiv \dim(M)-1$.
Note that $\square$ is precisely the Lichnerowicz wave
operator~\cite{Lichnerowicz:1961}. 
Also the appearance of the quadratic
Casimir implies that the constant curvature algebra, 
though finitely presented, is infinite dimensional.

A disadvantage of totally symmetric tensors as compared
to differential forms is that in a given dimension $d\equiv n+1$,
the space of symmetric tensors is unrestricted, 
{\it i.e.} the spectrum of $\N$ is ${\mathbb N}\cup\{0\}$
as compared to the $2^{n+1}$ possible differential forms\footnote{Higher spin
gauge fields of mixed symmetry expressed by 
expanding Young tableaux in columns, rather
than rows, are studied 
in~\cite{Hull:2001iu,deMedeiros:2002ge,deMedeiros:2003dc}.}.
(Of course, for physical higher spin applications, this is
in fact an advantage.) Hence practical computations in general
necessitate arbitrary functions (power series) in the the operators
$\TR$ and $\G$. A calculus for such computations is provided
by our next result:

The first step is to enlarge the constant curvature
algebra by a certain square root of the Casimir
\be
\Tt\equiv-\sqrt{\Big(\frac n2-\frac12\Big)^2-\c}
\ee
and in addition we define $\Nt\equiv \N+\frac n2-\frac 12$.
This allows us to form the operator $\Nt+\Tt$ whose eigenstates
are $k$-fold trace-free tensors, namely
\be
\TR^k \varphi =0\neq \TR^{k-1}\varphi
\ \Longrightarrow \
(\Nt+\Tt) \varphi = 2k \varphi\, .
\ee
Then introducing
\be
\wt \DIV\equiv (\Nt-\Tt)\DIV-\GRAD\TR, 
\ee
and similarly for the formal adjoint $\wt \GRAD$, the constant 
curvature algebra is presented by the six relations
\bea
\TR\Nt=(\Nt+2)\TR\,, &&
\TR\wt \GRAD=\wt \GRAD\ \frac{\Nt-\Tt+4}{\Nt-\Tt+2}\ \TR\, ,\nn\\[3mm]
\G\TR&=&\Nt^2-\Tt^2=\TR\G-4\Nt-4\, ,\nn\\[4mm]
\wt\DIV \Tt=(\Tt-1)\wt\DIV\, ,&&
\wt\DIV \Nt=(\Nt+1)\wt\DIV\, ,\nn\\[-2mm] \nn
\eea
$$
\wt\DIV\wt\GRAD\ =\ \wt\GRAD\wt\DIV\frac{(\Nt-\Tt+2)\Tt^2}{(\Nt-\Tt)(\Tt^2-1)}
$$
\be
\hspace{3.7cm}
-\ 2\frac{(\square-\frac{(n-1)^2}{2}+2\Tt^2)(\Nt-\Tt+2)\Tt^2}{\Nt-\Tt}\, .
\label{casalg}
\ee
and their formal adjoints where all other products are commutative.
In particular, observe that the $sl(2,\Real)$ action on the pair
$(\wt\DIV,\wt\GRAD)$ is diagonal.
These relations provide a calculus for constant curvature algebra
computations in terms of rational functions of $(\Nt,\Tt)$.

As the first, of what we anticipate to be many, applications of this calculus
we present the massive higher spin action generating function on a constant
curvature manifold\footnote{Here $(x)_n$ is the Pochhammer
symbol and the projector $\Pi_{\TR}=1-\delta_\TR=\G\frac{1}{(\Nt+2)^2-\Tt^2}\TR$.}
\bea
S&=&\varphi\ \Big[\ {\ss\square \ +\ \mu^2\ - \ (\Nt+\frac12)^2
\ +\ 2\ (\Nt-\frac n2+\frac12)(\Nt+\frac n2-\frac12)
\ -\ \frac14(\wt\GRAD-\G\wt \DIV)
\frac{1}{\Nt^2}\wt\DIV\ \Big] }
\ {\textstyle \delta_\TR}\ \varphi\nn\\[2mm]
&+&\varphi\ \Big[\ \ss{
\frac{4\ \Tt(\Tt-\frac12)}{(\Nt+\Tt)(\Tt-1)}\ 
\left(\square+2\ [\Tt-\frac n2+\frac12]
[\Tt+\frac n2-\frac12]
-\frac14\wt\GRAD \frac{\Tt-1}{\Tt(\Tt-\frac32)(\Nt-\Tt+2)}\wt\DIV
\right)}
\nn\\[1mm]&&\qquad\qquad\qquad\qquad\quad
{\ss \ - \ \left(\mu^2-[\Tt+\frac12]^2\right)
(\Nt-\Tt-1)} \Big]
\ {\ss 
\frac{\left(\frac{\Nt-\Tt+1}2\right)_{\!\sss \frac{\Nt+\Tt-2}2} }
{\left(\frac12\right)_{\!\sss \frac{\Nt+\Tt-2}2 }}}\ {\textstyle\Pi_{\TR}}
\  \varphi\nn\\[2mm]
&+&\!\!\! \ \Big\{2\ \varphi {\ss \G} \Big[ - 
{\ss\wt\GRAD  \ \frac{\mu-\Tt
+\frac32}{(\Nt+\Tt+2)}
\ +\  \G \
\frac{(\mu+\Tt-\frac32)(\Nt-\Tt+5)}{(\Nt-\Tt+2)(\Nt+\Tt+1)(\Nt+\Tt+4)}\wt\DIV\Big]}
\nn\\[1mm]&&
+\ \ \chi \, \ \Big[\ 
{\ss
\frac{4\ \Tt(\Tt-\frac12)}{\Tt-1}\left(\square+2
[\Tt-\frac n2+\frac12][\Tt+\frac n2-\frac12]
-\frac14\wt\GRAD
\frac{\Tt-1}{\Tt(\Tt-\frac32)(\Nt-\Tt+2)}
\wt\DIV
\right)}
\nn\\[1mm]&&\qquad
{\ss
\ -\  \left(
\mu^2-[\Tt+\frac12]^2\right)
(\Nt-\Tt+2)(\Nt+\Tt+3)}
\ \Big]
{\ss 
\frac{\Nt-\Tt+2}{\Nt+\Tt+1} } \Big\}
{\ss
\frac{\left(\frac{\Nt-\Tt+5}2\right)_{\!\sss\frac{\Nt+\Tt}2}}
{\left(\frac12\right)_{\!\sss\frac{\Nt+\Tt}2}} }
\ \chi\, .
\eea
Specializing to $(\N-s)\varphi=(\N-s+3)\chi$, this action describes
a spin~$s$ field of mass $m^2=-\mu^2+\Big(s+\frac n2-2\Big)^2$ in terms of a 
pair of unconstrained fields $(\varphi,\chi)$. 
Massless and partially
massless 
theories~\cite{Deser:2001us,Deser:2001pe,Deser:2001wx,Deser:2001xr,Deser:2003gw,Deser:2004ji} 
appear at tuned values of $m^2$ for which this 
action enjoys gauge invariances. A complete explanation may be found in 
Section~\ref{part}. Also, in the flat space limit, this result
provides a generating function for the original massive higher spin
actions of~\cite{Singh:1974qz}.

The remainder of the Article is organized thus; in Section~\ref{cca}
we define and describe the constant curvature algebra in detail.
These results are new. The following Section~\ref{zeromass} applies this
algebra to the massless action for higher spins in constant curvature
backgrounds, a subject studied already in detail in~\cite{Brink:2000ag,Vasiliev:2000rn}.
In Section~\ref{radial} we summarize the radial reduction technique
for reducing from Minkowski to constant curvature theories
developed in~\cite{Biswas:2002nk}. 
The combination of this reduction technique and our
constant curvature algebra allows us to solve the problem of writing 
a generating function for massive, constant curvature theories
of arbitrary higher spins in Section~\ref{gmass}. 
Some results exist already for these
theories\footnote{A light cone formulation of totally
symmetric Anti de Sitter higher
spins was given in~\cite{Metsaev:2003cu}.}~\cite{Zinoviev:2001dt}, 
but our algebraic generating function formalism, is
far more compact and suitable for generalizations, including perhaps
interactions. In Section~\ref{part}, we analyze partially massless
gauge symmetries of the massive action in terms of degeneracies of
the flat, massless gauge transformations in one dimension higher.
This result is also new. The fermionic version of the constant 
curvature algebra appears in Section~\ref{app1} in terms of the
corresponding superalgebra. We also write massless constant curvature
higher spin actions in terms of this algebra. 
Our reformulation of all higher spins in terms of a single scalar
field living on the total space of the cotangent bundle $T^*M$
is given in Section~\ref{total}.
Our conclusions are in
Section~\ref{conc}. In the Appendices, various aspects tangential to
the main developments are given. These include a new on-shell
formulation of partially massless higher spins in terms
of residual gauge transformations, as well as further details on 
the constant curvature algebra, in particular its interpretation
in terms of a non-commutative harmonic oscillator.

\section{Constant Curvature Algebra}

\label{cca}

Let $M$ be an $n+1$ dimensional constant curvature manifold.
Its Riemann curvature is
\be
R_{\m\n\r\s}=-\frac{2\L}{n}\,g_{\m[\r}g_{\s]\n}\ , 
\ee
where the parameter $\L$ is the cosmological constant, positive in de Sitter  
and negative in Anti de Sitter space. 
Throughout we will denote $n\equiv d-1$ and work in
units\footnote{This 
choice is germane to de Sitter space, however, upon
reinstating $\Lambda$ by dimensional 
analysis, all our formul\ae apply to Anti de Sitter as well.}
$\Lambda=n$.
The actions of commutators of covariant 
derivatives are therefore summarized by the vector-spinor example\footnote{Our 
metric is ``mostly plus'',
Dirac matrices are ``mostly hermitean'' and the Dirac conjugate is
$\ol \psi\equiv \psi^\dagger i\g^0$. 
We denote (anti)symmetrization with unit weight by round
(resp. square) brackets. Antisymmetrized products of Dirac matrices
are given by $\g^{\m_1\ldots\m_n}\equiv\g^{[\m_1}\cdots\g^{\m_n]}$.}
\be
[D_\m,D_\n]\,\psi_\r=2g_{\r[\m}\psi_{\n]}
+\frac12\g_{\m\n}\psi_\r\, .\label{covder}
\ee

Let $\odot T^*M$ be the symmetric tensor bundle whose 
sections are expressed in terms of tensor fields
\be
\odot T^*M\ni \Phi=\vphi_{\mu_1\ldots\mu_s}\ dx^{\mu_1}\ldots dx^{\mu_s}\, .
\ee
Note, there is no restriction to elements of definite index content, so
sums and products of elements with differing values of $s$ are allowed.
Here the product of differentials $dx^\mu$ stands for the symmetric 
tensor product\footnote{Or more correctly, the Cartan product, see~\cite{Eastwood:2004}.} 
\be
dx^{\mu_1}\ldots dx^{\mu_s}\equiv\frac{1}{s!}\sum_{\sigma}
dx^{\mu_{\sigma(1)}}\otimes\cdots\otimes dx^{\mu_{\sigma(s)}}\, .
\ee
In this way, we can regard symmetric tensors simply as {\it functions}
of the commuting differential $dx^\mu$.
We define various operations on symmetric 
tensors\footnote{This notational device has been considered before 
in~\cite{Manvelyan:2004ii}. In that notation
the vector $a^\mu$ should be regarded as a
coordinate differential $dx^\mu$.}. Firstly
\be
\d_\mu:\odot T^*M\rightarrow \odot T^*M
\ee
where
\be
\d_\mu dx^\nu = \delta^\nu_\mu + dx^\nu \d_\mu\, .
\ee
The coordinate vector $\d_\mu$ 
returns zero when acting all the way to the right
and obeys the Leibnitz rule on products of differentials. For example,
\be
\d_\mu\ dx^\nu dx^\rho=2\ \delta^{(\nu}_\mu dx^{\rho)}\, .
\label{del}
\ee
Importantly, the operation $\d_\mu$ is fiberwise and does
not act on the field-valued coefficients $\vphi_{\mu_1\ldots\mu_s}$.
The other operator we need is the covariant derivative
\be
D_\mu:\odot T^*M\rightarrow \odot T^*M
\ee
which acts only on the fields, not differentials
\be
D_\mu \Big[\vphi_{\mu_1\ldots\mu_s} dx^{\mu_1}\ldots dx^{\mu_s}\Big]
=D_\mu \vphi_{\mu_1\ldots\mu_s}\  dx^{\mu_1}\ldots dx^{\mu_s}\, .
\ee
To achieve a completely index free notation we define a slew of additional
operators built from the above ingredients:
\begin{itemize}
\item{\it Index:} \be {\N}\equiv dx^\mu \d_\mu \, .\ee
This operator counts indices. Its eigenvalues are simply
the number of indices and eigenspaces are tensors of definite
index type. For example, \be {\N}\  dx^{\mu_1}\ldots dx^{\mu_s}=
s\ dx^{\mu_1}\ldots dx^{\mu_s}\, .\ee
\item{\it Trace:} \be \TR\equiv g^{\mu\nu}\d_\mu\d_\nu\, .\ee
This is the operation of tracing over a single pair of indices. For example,
\be \TR\ \vphi_{\mu_1\ldots\mu_s} dx^{\mu_1}\ldots dx^{\mu_s}=
s(s-1)\ \vphi^\mu{}_{\mu\mu_3\ldots\mu_s}dx^{\mu_3}\ldots dx^{\mu_s}\, .\ee 
The kernel of this operator is the set of trace-free tensors
and eigenvectors of 
$\N$ with eigenvalue $0,1$ ({\it i.e.} scalar and vector fields).
\item{\it Metric:} \be \G\equiv g_{\mu\nu} dx^\mu dx^\nu\, .\ee
Forms the new tensor obtained by multiplying by the metric and symmetrizing, 
\be \G\ \vphi_{\mu_1\ldots\mu_s} dx^{\mu_1}\ldots dx^{\mu_s}=
 g_{\mu_1\mu_2}\vphi_{\mu_3\ldots\mu_{s+2}}dx^{\mu_1}\ldots dx^{\mu_{s+2}}
\, .\ee
\item{\it Divergence:} \be \DIV\equiv \d_\mu D^\mu\, . \ee
As suggested by its name
\be \DIV\ \vphi_{\mu_1\ldots\mu_s} dx^{\mu_1}\ldots dx^{\mu_s}=
s D_{\mu}\vphi^\mu{}_{\mu_2\ldots\mu_{s}}dx^{\mu_1}\ldots dx^{\mu_{s}}
\, .\ee
The kernel of $\DIV$ is the same as $\N$, {\it i.e.} scalar
fields.
\item{\it Gradient:} \be \GRAD\equiv dx^\mu D_\mu\, .\ee
This is the usual gradient operation 
\be\GRAD\ \vphi_{\mu_1\ldots\mu_s} dx^{\mu_1}\ldots dx^{\mu_s}=
 D_{\mu_1}\vphi_{\mu_2\ldots\mu_{s+1}}dx^{\mu_1}\ldots dx^{\mu_{s+1}}
\, .\ee
\item{\it Laplacian:} \be \Delta\equiv D^\mu D_\mu\, .\ee
An important relation in constant curvature spaces is
\be
\Delta=\DIV\ \GRAD -\GRAD\ \DIV
+\N(\N+n-1)-\G\TR\, .
\ee
The Laplacian is related to the commutator, rather than anticommutator,
of the divergence and gradient appearing in the theory of differential forms, because
we work with {\it symmetric} tensors. Moreover, in the flat space
limit the Laplacian coincides with the commutator of the divergence and
gradient.
\end{itemize}

\subsection{Algebra of Operations}

Much mileage is gained by computing the algebra of the operators
$\G$, $\GRAD$, $\N$, $\DIV$ and $\TR$. This is a simple calculation
whose results are simplified by noting the following
\begin{enumerate}
\item The operator
\be
\c=\G\TR-\N(\N+n-1)\label{casimir}
\ee
commutes with $\G$, $\TR$ and $\N$. In fact it is the quadratic
Casimir of an $sl(2)$ Lie algebra formed by this triplet.
\item The differential operator
\be
\square=\Delta+\c=[\DIV,\GRAD]+2\c\, ,
\ee
is central and is precisely the wave operator
introduced some time ago by Lichnerowicz~\cite{Lichnerowicz:1961}.
\end{enumerate}

\begin{figure}
\begin{center}
\epsfig{file=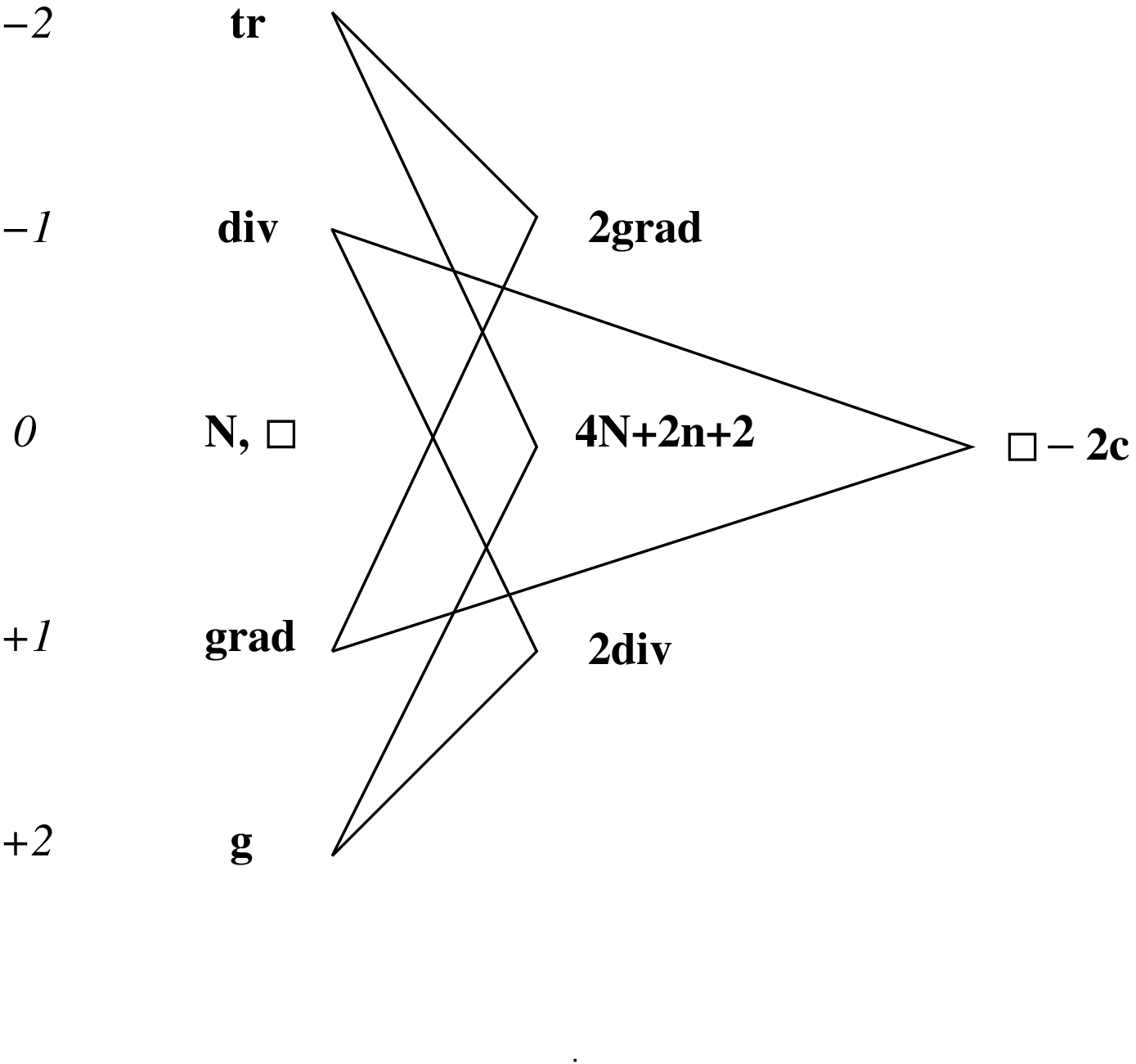,height=8cm}
\end{center}
\caption{The algebra of operators on symmetric tensors living on
$n+1$ dimensional constant curvature manifolds. The left-hand column
enumerates the weight with respect to the index operator $\N$.
The two 
rightmost columns are the results for non-trivial/vanishing commutators.
\label{commutators}}
\end{figure}

Without further ado, we present the algebra of operators on 
symmetric tensors
$$
[\TR,\G]=4\N+2n+2\,,\quad  [\TR,\GRAD]=2\DIV\,,\quad [\DIV,\G]=2\GRAD\, ,
$$
\be
[\DIV,\GRAD]=\square-2\c\, .\label{alg}
\ee
Commutators with the index operator $\N$ are all of the form
$[\N,{\cal O}]={\rm wt}_{\cal O}. {\cal O}$ where the weights are 
tabulated below
\be
\begin{array}{c|cccccc}
{\cal O}&\TR&\DIV&\N&\square&\GRAD&\G\\\hline
{\rm wt}_{\cal O}&-2&-1&0&0&1&2
\end{array}
\ee
All other commutators vanish and the algebra is neatly presented
by the diagram in Figure~\ref{commutators}. We will refer to this
algebra as the ``constant curvature algebra'' and its
flat space limit as the ``flat space algebra''. Finally, it is important
to note that although the flat space
algebra is a Lie algebra, its constant curvature deformation
by the Casimir $\c$, living in the universal enveloping algebra
of $sl(2,\Real)$, is not. The operators $\DIV$ and $\GRAD$
do {\it not} commute with $\c$ but rather higher commutators produce
an infinite sequence of additional operators. We stress that the
consistency of the algebra is, however, guaranteed thanks to its 
explicit representation acting on symmetric tensors.

\subsection{Analysis of the Operator Algebra}

Our first remark is that $\TR=-2 e_+$, $\G=2 e_-$ 
and $\N=-h-n/2-1/2$
generate an $sl(2,\Real)$ Lie algebra
\be
[h,e_\pm]=\pm 2e_\pm\, ,\qquad [e_+,e_-]=h\, .
\ee
The quadratic Casimir $\c=-h^2-2\{e_+,e_-\}+\frac14(n+1)(n-3)$
labels irreducible representations. Note that there is a natural adjoint 
operation 
\be
\TR^\dagger=\G\, ,\qquad \DIV^\dagger=-\GRAD\, ,\qquad N^\dagger=N\, ,
\qquad\square^\dagger=\square\, .\label{dagger}
\ee
Therefore, unitary $sl(2)$ irreducible representations are highest weight
where the highest weight state $\Phi$ obeys
\be
\N \Phi=s\Phi\, ,\qquad\TR\Phi=0\, ,
\ee
{\it i.e.} an $s$-index traceless symmetric tensor. The representation
is spanned by states $\{\Phi,\G\Phi, \G^2\Phi,...\}$ with $N$-eigenvalues 
$s,s+2,s+4,....$. This representation is infinite dimensional, as must be
on general grounds, and all states are eigenvectors of the Casimir $\c$ with 
eigenvalue $-s(s+n-1)$. Indeed, it is the well-known discrete series
representation of $sl(2,\Real)$.

Examining Figure~\ref{commutators}, we observe that the doublet
$\{\DIV,\GRAD\}$ is simply the fundamental representation of $sl(2,\Real)$.
Regarding the doublet as coordinates for the plane $\Real^2$, the triplet of 
operators $\{\TR,\N,\G\}$ act 
as infinitesimal generators of unimodular changes of basis.
However, since $\DIV$ and $\GRAD$ fail to commute, what we really have is 
a non-commutative plane. In flat space, we have only the central
deformation\footnote{In flat space, $\square$ becomes the usual
d'Alembertian.} $\square$ (on dimensional grounds the coefficient of
$-2c$ 
in~\eqn{alg} is $\Lambda/n\rightarrow 0$), while for constant curvature
manifolds there is an additional deformation by the Casimir $\c$ which is 
central in the $sl(2,\Real)$ subalgebra only. Finally, note that setting 
$\square-2c=0$, the algebra is a Lie algebra, closed under commutation
and is simply the $(2,1)$-parabolic of $sl(3,\Real)$ modded
out by its connected center. 

There is an an interesting relationship between the constant
curvature algebra and the harmonic oscillator. Also, the constant
curvature algebra in $n+1$ dimensions can be embedded in its flat
space counterpart in one dimension higher. Since these aspects
are tangential to our main story, they are presented in Appendix~\ref{harm}.
The computations we encounter in this Article involve functions
of arbitrarily high powers of the operators $\G$ and $\TR$ (although
typically the overall weight $N$ remains small). There are two
approaches to handle this difficulty. The first is choose a standard
ordering, for example normal ordering with $\G$'s to the left and $\TR$'s
to the right. Although this method will be useful at times, the
second method is far more powerful: namely to maximally commute
operators $\G$ and $\TR$ so that they are adjacent and then 
relate their product $\G\TR$ to the Casimir $\c$. We detail these
two techniques in the following Sections.

\subsubsection{Normal Ordering and Hadamard Products}
\label{hadamard}

For later use, we define a normal ordering for the constant curvature
and flat space algebras. Our convention is to order operators with
respect to their index operator $\N$ weights, {\it
i.e.}
\be
\G > \GRAD > \N,\square > \DIV>\TR\, .
\ee
We denote normal ordering as usual by $:\ldots:$, so for example,
\be
\TR\G\ =\ :\G\TR:+4\N+2n+2\, .
\ee
In addition, for later reference we will also need to compute
normal ordered expressions of the following type\footnote{Note 
that typically $f$ and $g$ will be even functions so that
no fractional powers of the operators $\G$ and $\TR$ appear.}:
\be
f(\partial\sqrt{\G}) :G(\G,\GRAD,\N,\DIV,\TR): g(du\sqrt{\TR})\, .
\ee
Here $[\partial,du]=1$ and $\partial$ vanishes when pushed all the
way to the right (as does $du$ to the left).
To solve this problem, note that only terms with equal numbers
of $\partial$ and $du$'s contribute so if $f(x)=\sum_n f_n x^n$
and $g(x)=\sum_n g_n x^n$ we obtain
\be
f(\partial\sqrt{\G}) :G: g(du\sqrt{\TR})
\ =\ :(f*g)(\sqrt{\G\TR}) \ G:\, ,
\ee
where the generalized Hadamard product
\be
(f*g)(x)\equiv\sum_n n! f_n g_n x^n\, . 
\ee

\subsubsection{Casimir Basis}

For many applications, we deal with operators
involving high powers of $\G$ and $\TR$.
These can be handled by the normal ordering
prescription given above. A more potent
technique is to express the product $\G\TR$
in terms of the Casimir $\c$ and $\N$ so that one
deals simply with functions thereof:

To begin with, introduce operators
\bea
\Nt&\equiv&\N+\frac n2-\frac12\, .\nn\\[1mm]
\Tt&\equiv&-\sqrt{\frac{(n-1)^2}{4}-\c}\, .
\eea
Here we are enlarging the constant curvature algebra 
by the above square root of the $sl(2)$ Casimir.
In fact, we will even employ rational functions of 
the operators $\Nt$, $\Tt$. This maneuver is extremely
useful because the combination $\Nt+\Tt$
measures how many traces annihilate a symmetric tensor
\be
(\Nt+\Tt) \G^k \xi=2k\G^k \xi\, \mbox{ when }\TR\xi=0\, .
\ee
Note also that
\be
\G\TR=(\Nt+\Tt)(\Nt-\Tt)=\TR\G-4\Nt-4\, .
\ee
and $\Tt$ commutes with $\G$, $\N$ and $\TR$ while
$\c=-(\Tt+\frac n2-\frac12)(\Tt-\frac n2+\frac12)$. Hence
\be
:(\G\TR)^k\!:\ =\G^k\TR^k=4^k\ 
\Big(\!-\frac{\Nt+\Tt}{2}\Big)_{\!k} \
\Big(\frac{\Tt-\Nt}{2}\Big)_{\!k}\, ,
\ee
where the Pochhammer symbol $(x)_k\equiv x(x+1)\cdots(x+k-1)$.
This observation allows us to convert normal ordered expressions 
to ones involving the Casimir and Index operators only. The archetypal example
is
\be
:\cosh\sqrt{\G\TR}:\ = 
\sum\frac{\G^k\TR^k}{(2k)!}
=\frac{\Big(\frac{1+{\cal N-T}}{2}\Big)_{\!\frac{\cal N+T}{2}}}
{\Big(\frac{1}{2} \Big)_{\!\frac{\cal N+T}{2}}}\, ,
\ee
where we note that the eigenvalues of
$\frac{\cal N+T}{2}$ are always non-negative 
integers.

\subsubsection*{Casimir Algebra}

It is most useful to introduce trace and spin modifications of the operators
$\DIV$ and $\GRAD$
\be
\wt\DIV\equiv(\Nt-\Tt)\DIV-\GRAD\TR\, ,\qquad
\wt\GRAD\equiv\GRAD(\Nt-\Tt)-\G\DIV\, ,\label{tilde}
\ee
which have simple commutation relations with the operator $\Tt$
\be
\left[\Tt,\left(\begin{array}{c}\wt\DIV\\[1mm]\wt\GRAD\end{array}\right)
\right]=\left(\begin{array}{c}\wt\DIV\\[1mm]-\wt\GRAD\end{array}\right)\, .
\ee
We may invert the relations~\eqn{tilde} by solving a $2\times 2$ 
operator-valued matrix problem and find
\bea
\DIV&=&-\frac{1}{2}\ \Big(\frac{1}\Tt\ \wt\DIV+
\wt\GRAD\TR\ \frac{1}{\Tt(\Nt-\Tt)}\Big)\, 
,\nn\\
\GRAD&=&-\frac{1}{2}\ \Big(\wt\GRAD\ \frac{1}{\Tt}\
+\frac{1}{\Tt(\Nt-\Tt)}
\G\wt\DIV\Big)
\, .
\label{undo}
\eea
Using these relations it is now not difficult to reproduce the 
algebra~\eqn{casalg} quoted in the Introduction.

Let us emphasize what we have achieved: The infinite dimensional
Lie algebra~\eqn{alg} has been rephrased as a finitely presented
algebra in seven generators.
Needless to say the
Casimir basis is highly advantageous for calculations.

We end this Section with a warning and two new operators:
Computations involving rational functions of $\Nt$ and $\Tt$
risk encountering singular denominators. The most important 
example is the operator
\be
{\bf \Pi_{\TR}}\equiv\G\ \frac{1}{(\Nt+2)^2-\Tt}\ \TR\, .
\ee
One might think that this operator were unity by commuting
the denominator left past $\G$ (which shifts $\Nt$ by $-2$)
and then using the relation $\G\TR=\Nt^2-\Tt^2$ to conclude
that $(\Nt^2-\Tt^2)^{-1}\G\TR=1$. This is clearly false. The operator 
${\bf \Pi_{\TR}}$ is actually a projector, returning zero on traceless tensors
and unity otherwise. We also define its complement ${\bf \delta_{\TR}}\equiv 1-
{\bf \Pi_{\TR}}$
which projects onto traceless tensors.

\section{Generating Function for Massless Higher Spin Actions}

\label{zeromass}

Our first physics application of the constant curvature and 
flat space algebras is to write a generating function for massless
higher spins. The massless theories are very well known 
and have rather simple actions, indeed our result amounts to 
little more than simply replacing the eigenvalue $s$ of the index operator
by the operator $\N$. The key point, however, is that we present
an action depending on fields whose spin need not not be specified
(and can even be indefinite) and which generates all massless
actions when evaluated on fields of definite spins.

The minimal covariant field content for massless higher spins 
is a doubly traceless field $\varphi$:
\be
\TR^2\varphi=0.
\ee
The correct physical degrees of freedom are ensured by the gauge
invariance
\be
\delta\varphi=\GRAD\xi\, ,\label{m=0gauge}
\ee
where the gauge parameter $\xi$ is tracefree:
\be
\TR\xi=0.
\ee
To study actions we need to introduce an inner product on the space
of symmetric tensor fields
\be
\langle \varphi,\chi\rangle = \int\sqrt{-g}\ \varphi^\dagger \chi\, .
\label{dual}
\ee
The adjoint operation $\dagger$ on operators was given in~\eqn{dagger}
and acts on fields via
\be
{}[\varphi^*_{\mu_1\ldots\mu_s}dx^{\mu_1}\cdots dx^{\mu_s}]^\dagger=
\varphi^{\mu_1\ldots\mu_s}
\d_{\mu_1}\cdots \d_{\mu_s}\, .
\ee
The operators $\d_{\mu_i}$ must then be passed all the way to the right
where they return zero. It is extremely important to note that this dualization is metric dependent
since $\varphi^{\m_1\ldots\mu_s}=g^{\mu_1\nu_1}\cdots g^{\mu_s\nu_s}
\varphi_{\nu_1\ldots\nu_s}$.
We will often use the shorthand
notation $\langle \varphi,\chi\rangle\equiv \int \varphi\chi$, so for example
when $\varphi=\varphi_\mu dx^\mu$ and $\chi=\chi_\mu dx^\mu$ we have
\be
\int \varphi \chi=\int \sqrt{-g} \ \varphi_\mu^* \chi^\mu\, .
\ee

The action for massless fields is then uniquely determined (assuming
quadratic derivatives) by the gauge invariance~\eqn{m=0gauge} and 
reads\footnote{Integrating by parts, we may also write
$$
S=-\int\Big\{|\GRAD\varphi|^2-2|\DIV\varphi|^2+(\DIV\varphi)^\dagger\GRAD\TR\varphi
-\frac12|\GRAD\TR\varphi|^2+\frac14|\DIV\TR\varphi|^2$$
$$-2\varphi^\dagger(\N-1)(\N+n-2)\varphi
+\frac12(\TR\varphi)^\dagger(2\N(\N+n)+n)\TR\varphi\Big\}\, .
$$.}
\bea
S&=&\int \varphi \Big\{
\square+2(\N-1)(\N+n-2)-\GRAD\DIV+\frac12[\GRAD^2\TR+\G\DIV^2]
\nn\\&&\qquad
-\frac12\G[\square+2\N(\N+n)+n+\frac12\GRAD\DIV]\TR
\Big\} \varphi\ \equiv \ \int \varphi\,  G \varphi\, .\nn\\\label{m0L}
\eea
The half integer spin generalization of this result is given
in Section~\ref{fermions}. Its voracity is easily verified by computing
the Bianchi Identity
\be
\DIV G = \G X  \,\equiv 0 \mbox{ mod } \G,
\ee
corresponding to the gauge invariance~\eqn{m=0gauge}.

\section{Radial Dimensional Reduction}

\label{radial}

The $d\equiv n+1$ dimensional 
de Sitter spacetime\footnote{All computations in this Section
generalize trivially to Anti de Sitter (see~\cite{Biswas:2002nk} 
for details). We
concentrate on de Sitter for definiteness only.} 
is most simply described by pulling back the
flat metric 
\be
ds^2=-dZ^2 + \sum_{i=1}^d \Big( dX^i\Big)^2
\ee 
in $d+1$ dimensions to the hyperboloid
\be
-Z^2 + \sum_{i=1}^d \Big( X^i\Big)^2 =\frac{n}{\Lambda}\, .
\ee
As usual, we set $\Lambda/n=1$, factors of which can 
be readily reinstated by examining dimensions.
In particular, choosing new Minkowski coordinates
$Z=e^u\xi^0$, $X^i=e^u\xi^i$ where $\xi^M\eta_{MN}\xi^N=1$
so that $e^{2u}=-Z^2+ \sum_{i=1}^d ( X^i)^2$,
we have $ds^2=e^{2u}( du^2 + d\xi^M \eta_{MN} d\xi^N)$ and in turn
\bea
ds^2= \exp(2u)\Big(du^2+ ds^2_{dS}(x)\Big).\label{flatmetric}
\eea
The de Sitter metric $ds^2_{dS}(x)$ in the second line appears 
upon choosing a parametrization of the hyperboloid $\xi^M=\xi^M(x^\mu)$.
In particular observe that the operator
$\partial_u\equiv\frac{\partial}{\partial u}$
is a homothetic Killing vector $[\partial_u,ds^2]=2ds^2$.
For this reason, a (rather elegant) radial dimensional reduction from flat
to de Sitter space with respect to
the conformal isometry generated by $\partial_u$ 
was suggested in~\cite{Biswas:2002nk}.

This reduction is easily understood by looking at the simplest
example of a massless scalar
\be
S=-\frac12 \int dZ d^dX\  \partial_M \Phi  \eta^{MN}  \partial_N \Phi\, .
\ee
In the coordinates $(u,x^\mu)$ the action is
\be
S=-\frac{1}{2}\int du\ e^{nu}
d^dx\sqrt{-g_{dS}}\ \Big\{\ 
\partial_\mu \Phi  g^{\mu\nu}_{dS} 
\partial_\nu \Phi+[\partial_u \Phi]^2\Big\}\, .
\ee
Crucially, we wish to make a Scherk--Schwarz reduction~\cite{Scherk:1979zr} with respect
to the log-radial coordinate $u\in(-\infty,\infty)$. 
Therefore
we demand that the measure of the action functional be simply $du$
which requires
the Weyl rescaling
\be
\Phi=e^{-nu/2}\vphi\, .\label{weyl0}
\ee 
We remark, that $n/2$ is of course the scaling dimension of the field $\Phi$.
The action now becomes
\be
S=-\frac{1}{2}\ 
\int du d^dx\sqrt{-g_{dS}}\ \Big\{\ 
\partial_\mu \vphi  g^{\mu\nu}_{dS} \partial_\nu \vphi+
 \Big[(\partial_u-\frac{n}{2})\vphi\Big]^2
\Big\}\, .
\ee
We may now Scherk--Schwarz reduce to the
sector of fixed log-radial momentum\footnote{It is most
important to note that $\int du f(u) g(u)=\frac{1}{2\pi}
\int dp_u f(-p_u) g(p_u)$ reflecting log-radial momentum conservation.} 
\be\partial_u=im\, .\ee
The reduced action reads
\be
S_{dS}=-\frac{1}{2}\int d^dx\sqrt{-g_{dS}}\ \Big\{\
\partial_\mu \vphi  g^{\mu\nu}_{dS} \partial_\nu \vphi
+[m^2+\frac{n^2}{4}]\ \vphi^2\Big\}\, .\label{redS}
\ee
A simple consistency check verifies that the reduced
equation of motion $(-\square_{dS}+m^2+\frac{n\Lambda}{4})\vphi=0$
obtained by varying the reduced action~\eqn{redS}
coincides with original $d+1$ dimensional field equation 
$-\square_{d+1} \Phi=0$ in this fixed momentum sector.
In the next Section we apply the radial dimensional reduction
technique to derive a generating function
for all massive and partially massless theories in de Sitter space.

\section{Generating Functions for Massive Actions}

\label{gmass}

An important observation is that constant curvature
massive and partially massless
systems in dimension $d$ are really just subsectors of the 
$d+1$ dimensional, flat massless theory.
In particular as we shall see, the partially massless theories
arise as degeneracies of the $d+1$ dimensional gauge variations.
It was first observed that massive higher spins could be obtained
by dimensional reductions in~\cite{Aragone:1988yx} 
(see~\cite{Bekaert:2003uc} for more recent
developments).  

Our starting point therefore is the 
massless higher spin system in $d+1$ flat dimensions
formulated in terms of a doubly traceless
field $\Phi$, 
\be\TTR^2\Phi=0\, ,\label{trtr}\ee subject to a gauge invariance
\be
\delta\Phi=\GGRAD \Xi\, , \label{Gauge}
\ee
with traceless gauge parameter \be\TTR\Xi=0.\label{tr}\ee
We distinguish $d+1$ dimensional flat operators from their
$d$ dimensional constant curvature counterparts by capitalization.

The first maneuver is to make Weyl rescalings mimicking the scalar
case~\eqn{weyl0} with the usual spin dependence of the
conformal weight\footnote{Note that although the appearance of 
$\N_{d+1}$ in the scaling dimensions
is canonical, it can be deduced by looking at the leading behavior
of the flat d'Alembertian $\square_{d+1}=\partial_u^2+[s+n/2]\partial_u+\cdots$
acting on a spin $s$ field. The Weyl rescalings quoted above precisely cancel
the terms linear in $\partial_u$.}:
\be
\Phi\equiv e^{(\N_{d+1}-n/2)u}\phi \, ,\qquad
\Xi\equiv e^{(\N_{d+1}-n/2+1)u}\xi\, .
\ee

Next we need to decompose the $d+1$ tensorial structure of
$\Phi$ and $\Xi$ into $d$ dimensional tensors.
Both the field $\phi$ and gauge parameter $\xi$ can be expanded
in a power series in the commuting differential $du$
\be
\phi=\vphi_0+du\, \vphi_1 +
\frac1{2!}du^2\, \vphi_2+\frac1{3!}du^3\, \vphi_3+
\cdots\, , \qquad
\xi=\xi_1+du\, \xi_2 +\cdots\, .\label{fieldexp} 
\ee
If $\phi$ is of definite index, {\it i.e.} $\N_{d+1}\phi=s\phi$, 
the fields $\varphi_t$ have definite index in one dimension lower
$\N \vphi_t=(s-t)\vphi_t$. In turn the gauge parameters
$\xi_t$ then have index $\N \xi_t=(s-t)\xi_t$.

A beauty of the dimensional reduction technique is that the 
towers of auxiliary fields required for massive covariant
actions appear naturally. Moreover, in our approach,
these towers of auxiliaries are packaged in a minimal
set of {\it unconstrained} covariant fields.

To solve the traceless and double traceless
conditions~\eqn{tr} and~\eqn{trtr} in terms of a minimal set of
unconstrained $d$ dimensional fields, observe that
$[\partial,\TR]=0$ so $\TTR \Xi=0$ $\Rightarrow$ $(\partial^2+\TR) \xi=0$
is the equation for a harmonic oscillator with solution
\be
\xi=\cos(du\sqrt{\TR})\xi_1+\frac{\sin(du\sqrt{\TR})}{\sqrt{\TR}}\xi_2\, .
\label{gauge}
\ee
We have chosen boundary conditions such that $\xi$ is determined in
terms of the $d$ dimensional unconstrained gauge parameters \
$\xi_1$ and $\xi_2$.
We can solve the double trace condition~\eqn{trtr} in exactly the same
way and find
\be
\phi\!=\!\cos(du\sqrt{\!\TR})\!\Big[\!
\vphi_0\!-\!\frac12du(\vphi_1\!\!+\!\frac{1}{\TR}\vphi_3)\!\Big]
\!+\!\frac{\sin(du\sqrt{\!\TR})}{2\sqrt{\TR}}\!\Big[\!
3\vphi_1\!+\!\frac{1}{\TR}\vphi_3\!+\!du(\TR\vphi_0\!+\!\vphi_2)\!\Big]\!  .
\label{field}
\ee
Note that despite the appearance of explicit inverse powers of $\TR$,
the above expressions are actually regular at $\TR=0$ and at any fixed
spin their expansions terminate.
As boundary conditions we have taken the leading terms of the
expansions displayed in~\eqn{fieldexp}. In particular this means 
that a massive spin~$s$ field is described by four
{\it unconstrained} fields $\varphi_{0,1,2,3}$ of spin $s,s-1,s-2,s-3$
modulo St\"uckelberg gauge transformations parameterized by
two unconstrained fields $\xi_1$ and $\xi_2$ with spins\footnote{
The higher spin St\"uckelberg formalism has been studied recently
in~\cite{Bianchi:2005ze}.} $s-1,s-2$.

\subsection{Radially Reduced Gauge Variations}

To compute the radially reduced gauge transformations of
the four independent, unconstrained, 
$d$ dimensional fields $\varphi_{0,1,2,3}$
we first need to radially reduce the operator $\GGRAD$.
This is a simple matter of spelling out the Christoffel 
symbols for the flat metric in log-radial coordinate~\eqn{flatmetric}s
and substituting these in $\GGRAD=du D_u + dx^\mu D_\mu$ acting on
an  arbitrary tensor field. 
This easy, but slightly tedious computation yields
\be
\GGRAD=\GRAD+\G\partial+du(\partial_u-2N-du\partial)\, ,
\ee
as quoted in Appendix~\ref{in_bed} where the embedding of the
 $d$ dimensional constant curvature algebra  in the $d+1$ dimensional
flat space algebra is
described in detail.
Hence, because
\be
\delta\phi=e^{-(\N_{d+1}-n/2)u}\ \GGRAD\  e^{(\N_{d+1}-n/2+1)u}\xi\, ,\ee
and the commutator $[N_{d+1},\GGRAD]=\GGRAD$, we obtain
\be
\delta\phi=\Big[\GGRAD+(\N_{d+1}-n/2)du\Big]\ \xi\, ,
\ee
where the second term on the right hand side appears because
$[\GGRAD,u]=du$.
To compute $\GGRAD\xi$ explicitly we note the following identities
\bea
[\G, f(\TR)]\quad&=&4\TR f''(\TR)-4f'(\TR)\ (\N+d/2)\, ,\\
{}[\GRAD,f(\TR)]&=&-2f'(\TR) \DIV\, ,\\
{}[\N,f(\TR)]\quad\!&=&-2\TR f'(\TR)\, .
\eea
So finally, orchestrating
these instruments, a straightforward 
computation based on~\eqn{gauge} and~\eqn{field}
gives the rather compact gauge variations
\bea
\delta\vphi_0&=&\;\ \ \GRAD\xi_1\ \ +\ \ \G\xi_2\label{GG1}\\
\delta\vphi_1&=&\;\ \ \GRAD\xi_2\ \ +\ \ (\d_u-[\N+1]-\frac
n2+2-\G\TR)\xi_1
\label{GG2}\\
\delta\vphi_2&=&-\GRAD\TR\xi_1+2(\d_u-[\N+2]-\frac n2+3-\frac12\G\TR)
                \xi_2\label{GG3}\\
\delta\vphi_3&=&-\GRAD\TR\xi_2-3(\d_u-[\N+3]-\frac n2+4-\frac13\G\TR)
                \TR\xi_1\, ,
\label{GG4}
\eea
of the unconstrained $d$ dimensional fields $\vphi_0$, $\vphi_1$,
$\vphi_2$ and $\vphi_3$ equivalent to that of the original 
$d+1$ dimensional field $\Phi$ in~\eqn{Gauge}.

\subsection{Analysis of Gauge Transformations}
\label{gaugeanaly}

The gauge transformations~\eqn{GG1}-\eqn{GG4} 
play a central r\^ole in what follows.
As discussed in~\cite{Dolan:2001ih,Deser:2003gw}, masses of bulk fields in de Sitter are
related to conformal weights of boundary fields living on an
initial Cauchy surface. Since we want to relate the operator $\partial_u$ 
to the mass by a Scherk--Schwarz reduction, we posit the relationship
\be
\partial_u=\Delta_s-\frac n2\, ,
\ee
to be verified by the results that follow,
where $\Delta_s$ obeys the spin~$s$ mass-conformal weight relation
\be
m^2=-\Delta_s(\Delta_s-n)+(s-2)(s-2+n)\, ,
\ee
discovered in~\cite{Deser:2003gw}. 
Of particular interest are partially massless
depth $t$ tunings of the conformal weight 
\be
\Delta_s=n+s-t-1=\partial_u+\frac n2\, .
\label{tunes}\ee
If we fix the spin of $\varphi_0$ to be $s$ then the coefficients of the
tracefree components of $\xi_1$, $\xi_2$ and $\TR\xi_1$ in~\eqn{GG2}-\eqn{GG4}
exactly 
reproduce the depth $t=1,2,3$ tuning relations, respectively.
However, the origin of the remaining tunings for $t=4,\ldots,s$
are for now, obscure. Note however, that at depth $t$ these are
governed by a traceless parameter $\overline\xi_t$ 
with index $\N\xi_t=(s-t)\overline\xi_t$
and mass tuning $(\partial_u+\Tt+\frac12)\overline\xi_t=0$. (Note that
traceless fields obey $\Nt=-\Tt$.)

Observe that when we Scherk--Schwarz identify $\partial_u$ with
a {\it generic} mass parameter $\mu$, the $\xi_1$, $\xi_2$ gauge invariances 
are just shifts so that $\varphi_2$, $\varphi_3$ and the tracefree part of
$\varphi_1$ can be gauged away\footnote{Actually, as discussed
below, it is more convenient to gauge away linear
combinations of $\varphi_2$, $\varphi_3$ and $\TR\varphi_0$, $\TR\varphi_1$.}. 
{\it I.e.}, these are St\"uckelberg
fields. The massive models can be formulated in terms of a minimal
covariant field content $\varphi\equiv \varphi_0$ 
and $\chi\equiv \frac{1}{\TR\G}\TR\varphi_1$.
However, for special values of the mass, the above transformations
develop singularities corresponding to partially massless theories:

Partially massless gauge theories are based on higher derivative gauge 
invariances. At depth $t$ these take the form
\be
\delta\vphi_0=(\GRAD^t+\cdots)\overline\xi_t\, ,
\label{cdots}
\ee
where the parameter $\overline\xi_t$ is traceless spin $s-t$.
An important geometric result would be
an explicit formula for ``$\cdots$'':
The trick is to find an appropriate field redefinition
after which the gauge transformations can be written
as a matrix with zeros everywhere save for its diagonal 
and next to diagonal entries. This allows us to readily
identify degeneracies of the gauge transformations at special 
values of the mass parameter $\mu$
corresponding to partially massless theories. At other generic values,
as discussed above, the gauge transformations are just St\"uckelberg
shift symmetries.
The required redefinition is
\be
\wt\vphi_2=\vphi_2+\frac{\Nt+\Tt}{\Nt+\Tt+2}
\TR\vphi_0\, ,\qquad
\wt\vphi_3=\vphi_3+\frac{\Nt+\Tt}{\Nt+\Tt+2}
\TR\vphi_1\, .\label{redef}
\ee
which leads to gauge transformations
\bea
\delta\wt\vphi_2&\!=\!&
-2\, \wt\GRAD\ \frac{\Tt-1}{\Tt((\Nt+2)^2-\Tt^2)}\TR\ \xi_1
\ +\ 2 \ \Big(\partial_u+\Tt+\frac12\Big)\ \xi_2\, ,\nn\\[3mm]
\delta\wt\vphi_3&\!=\!&
-2\, \wt\GRAD\ \frac{\Tt-1}{\Tt((\Nt+2)^2-\Tt^2)}\TR\ \xi_2
-2\ \frac{\Nt\!+\!\Tt\!+\!3}{\Nt\!+\!\Tt\!+\!2}
\Big(\partial_u+\Tt+\frac12\Big)\TR\ \xi_1\, .\nn\\
\label{stuck}
\eea 
To appreciate the significance of these formul\ae, 
expand the gauge parameters in traceless pieces
\bea
\xi_1&=&\xi_1^0+\G\xi_1^2+\G^2\xi_1^4+\cdots\, ,\nn\\[1mm]
\xi_2&=&\xi_2^0+\G\xi_2^2+\G^2\xi_2^4+\cdots\, ,\nn\\[1mm]
\eea
and introduce a column vector $(\xi_j)=
(\xi_1^0,\xi_2^0,\xi_1^2,\xi_2^2,\xi_1^4,\ldots)$
by identifying $\xi_{2k+1}$ with $\xi_1^{2k}$ and
$\xi_{2k+2}$ with $\xi_2^{2k}$. 
(In general, we denote the expansion of a symmetric tensor $X$ 
into tracefree pieces $X^{2k}$ by
$X=X^0+gX^2+g^2X^4+\cdots$.)
As mentioned above,
the fields $\varphi_1^0$, $\wt\varphi_2$ and $\wt\varphi_3$ 
are St\"uckelberg at generic values of the mass and can be algebraically gauged
away. However, expressing their gauge transformations
as a matrix acting on the column vector $(\xi_t)$, we find
\be
\delta
\left(\begin{array}{c}
\varphi_1^0\\[2mm]
\wt\varphi_2^0\\[2mm]
\wt\varphi_3^0\\[2mm]
\wt\varphi_2^2\\[2mm]
\vdots\\[2mm]
\end{array}\right)
 = 
\left(\begin{array}{ccccc}
  u_1 &  v_1&\\[2mm]
&    
  u_2 & v_2
\\[2mm]
&&     u_3  &
 v_3 \\[0mm]
&&&  u_4 &
 \ddots \\[1mm]
&&&& \ddots \\[2mm]
\end{array}\right)
 \left(\begin{array}{c}
\xi_1^0\\[2mm]
\xi_2^0\\[2mm]
\xi_1^2\\[2mm]
\xi_2^2\\[3mm]
\vdots\\[2mm]
\end{array}\right)\equiv M . (\xi_t)\, .\label{sGG}
\ee
where the diagonal entries are the tunings
\be
u_j\propto \partial_u+\Tt+\frac12.\label{93}
\ee
(Notice that by tracelessness we have $(\partial_u+\Tt+\frac12)\overline\xi_t
=(\partial_u-\frac n2-s+t+1)\overline\xi_t$ 
which agrees with~\eqn{tunes}.)
Moreover the next to leading diagonal entries are
simply $v_j\propto \wt\GRAD$. Hence, whenever
$\det M\neq 0$, {\it i.e.} in the case that the diagonal tunings 
$u_j$ do not vanish, the theory is described by a massive
action \be S(\varphi\equiv\varphi_0,\varphi_1\equiv 
g\chi,\wt\varphi_2=0,\wt\varphi_3=0)\, ,\label{zeroact}\ee 
with no gauge invariances
remaining. However, whenever some 
\be
u_t=0\, ,
\ee
the matrix $M$ is degenerate. In that case we can still choose $$(\xi_j^>)
=\left(\begin{array}{c}0 \\ \vdots \\ 0 \\ \xi_{t+1} \\ \xi_{t+2}
\\ \vdots\end{array}\right)$$ 
so as to gauge away
the corresponding lower components of $\wt \varphi_2$, $\wt \varphi_3$.
We can then arrange for the action~\eqn{zeroact}, with the same 
field content as in the massive case, to be gauge invariant
by solving for a zero mode 
\be(\xi_t^\leq)=
\left(\begin{array}{c}\prod_{i=1}^{t-1}\Big(-u_{i}^{-1}v_{i}\Big).\,
\overline\xi_t \\[2mm]\vdots\\[2mm]
-u_{t-1}^{-1}v_{t-1}\overline\xi_t
\\[2mm]\overline\xi_t\\[2mm] 0\\[2mm] 0 \\[2mm]\vdots
\end{array}\right)\, ,\label{zeromode}\ee
of $M$ with a single, independent, traceless gauge
parameter $\overline\xi_t$. 
This works because the upper components of $\wt \varphi_2$, $\wt\varphi_3$
and $\varphi_1^0$ do not transform under this gauge transformation.
Specializing to spin~$s$, there are precisely $s$ such degeneracies
which correspond to partial gauge invariances of the generically
massive action~\eqn{zeroact}. We analyze these transformations and theories 
in depth in Section~\ref{part}. 
(In particular, the zero mode~\eqn{zeromode} 
provides the promised ``$\cdots$'' of~\eqn{cdots}.)
A pressing task is the computation of the $d$ dimensional action
$S(\varphi_0,\varphi_1,\wt\varphi_2,\wt\varphi_3)$ and its
truncation~\eqn{zeroact}. The latter action describes both massive and
partially massless theories. We 
first illustrate the above ideas for the case of spin~2.

\subsection{Example: Spin~2}

Let us denote $\vphi_0\equiv h$, $\vphi_1\equiv A$ 
and $\vphi_2=\wt\vphi_2\equiv\sigma$, 
being a symmetric tensor, vector and scalar, respectively.
The gauge symmetries~\eqn{GG1},~\eqn{GG2} and~\eqn{stuck} read
\bea
\delta h&=&\phantom{-}\,\GRAD\, \xi_1\ \ +\ \ \ \G\ \xi_2\nn\\[2mm]
\delta A&=&\phantom{-}\,\GRAD\, \xi_2\ \ +\ \ (\d_u-\frac n2)\ \xi_1\nn\\
\delta \sigma&=&-\GRAD\TR\xi_1+2\ (\d_u-\frac n2+1)\xi_2\, .
\eea
The massive theory is obtained by a Scherk--Schwarz reduction
\be
\d_u^2+m^2=\frac{n^2}4\, .\label{mass}
\ee
In this case both $A$ and $\sigma$ are St\"uckelberg fields
that can be algebraically gauged away by choices of $\xi_1$ and $\xi_2$.
However, if we examine more carefully the matrix of gauge transformations
for the candidate St\"uckelberg fields, {\it i.e.}, equation~\eqn{sGG},
we have
\be
\delta
\left(
\begin{array}{c}
A\\[3mm] \sigma
\end{array}
\right)
\ = \
\left(
\begin{array}{cc}
\d_u-\frac n2&\GRAD\\[3mm] &2\, (\d_u-\frac n2 +1)
\end{array}
\right)
\left(
\begin{array}{c}
\xi_1\\[3mm] \xi_2
\end{array}
\right)\, ,
\ee 
which is degenerate whenever
\be
\d_u=\frac n2 \ \mbox{ or }\  \d_u=\frac n2-1\, .
\ee
Inserting these relations in~\eqn{mass} (and reinstating the cosmological
constant) gives masses
\be
m^2=0 \ \mbox{ or }\  m^2=\frac{n-1}{n} \ \Lambda \, .
\ee
(In dimension~4, the latter is the well-known $m^2=2\Lambda/3$ tuning
for partially massless spin~2 fields~\cite{Deser:1983tm,Deser:1983mm}.) 
Hence, the zero eigenvectors
in these two cases
\be
\left(
\begin{array}{c}
\overline\xi_1\\[3mm] 0
\end{array}
\right)
\ \mbox{ or }\ 
\left(
\begin{array}{c}
\GRAD\overline \xi_2\\[3mm] \overline\xi_2
\end{array}
\right)
\ee
correspond to gauge invariances 
\be
\delta h=\GRAD\overline\xi_1
\ \mbox{ or }\ 
\delta h=\GRAD^2\overline\xi_2+\G\overline\xi_2\, ,
\ee
of the theory after setting $A=\sigma=0$.
The first invariance is the linearized diffeomorphism invariance
of massless spin~2 fields (gravitons), 
while the second is the scalar gauge
invariance of the partially massless spin~2 theory. 

\subsection{Radially Reduced Action}

Our major result is the massive, constant curvature, 
higher spin action obtained by radial reduction.
The computation proceeds in the following steps:
\begin{enumerate}
\item[(0)] The starting point is the {\it massless} action in $d+1$
flat dimensions\footnote{The flat massless actions
date back to the work of~\cite{Fronsdal:1978rb,Fang:1978wz,Curtright:1979uz}.} 
(obtained by letting $\Lambda\rightarrow 0$ in~\eqn{m0L})
\be
S=\int\Phi\, \{{\ss
\square_{d+1}-\GGRAD\DDIV+\frac12[\GGRAD^2\TTR+\GG\DDIV^2]
-\frac12\GG[\square_{d+1}+\frac12\GGRAD\DDIV]\TTR
}\}\, \Phi\,  .
\ee
\item[(i)] We Weyl rescale the $d+1$ dimensional, real,  massless field by
$$\Phi\equiv e^{(\N_{d+1}-n/2)u}\phi\, ,$$ as discussed above.  
All exponentials of $u$ and therefore explicit $u$-dependence cancels
when we remember the $u$ dependence of the metric determinant, the 
dualization defined in~\eqn{dual} and the exponentials
appearing in the embedded algebra~\eqn{algembed}. 
However, to achieve this end, we must 
first pass the Weyl exponentials through the operators $\DDIV$ and $\GGRAD$ 
according to
$$
\GGRAD \ e^{(\N_{d+1}+\alpha)u}=
 e^{(\N_{d+1}+\alpha-1)u} \Big[\GGRAD+(\N_{d+1}+\alpha-1)du\Big]\, ,\qquad
$$
\be
\DDIV \ e^{(\N_{d+1}+\alpha)u}=
 e^{(\N_{d+1}+\alpha+1)u} \Big[\DDIV+(\N_{d+1}+\alpha+1)\partial\Big]
\, .
\ee
We may also now express operators of the $d+1$ dimensional 
flat space algebra in terms of $d$ dimensional constant curvature ones.
(The factors $\exp\{\pm 2u\}$ appearing in~\eqn{algembed} 
are handled as above.)
\item[(ii)]
The action now takes the form
\be
S=\Big[\int_{-\infty}^\infty\! du\Big]\int\phi\  \widetilde
G(\G,\GRAD,\N,\square,\DIV,\TR;du,\partial,\partial_u)\, \phi\, ,\ee
with {\it no} explicit $u$-dependence and the integration measure is
the $d$ dimensional constant curvature one. Hence the operator
$\partial_u$ is by now central. The one dimensional Heisenberg
algebra
$(du,\partial)$ 
commutes with the $d$ dimensional constant curvature algebra. It is important to
remember that the
field $\phi$ and its dual depend on $du$ and $\partial$ as indicated
in~\eqn{field}. 
\item[(iii)] Next we must perform the 
$(\partial,du)$ algebra in terms of the generalized Hadamard product
defined in Section~\ref{hadamard}. Noting that $\partial f(du\sqrt{\TR})
=\sqrt{\TR}f'(du\sqrt{\TR})$, this problem reduces to the following
Hadamard products
$$
(\cos*\cos)(x)=\cosh(x)\, ,\qquad
(\sin*\sin)(x)=\sinh(x)\, ,
$$
\be
(\cos*\sin)(x)=0=
(\sin*\cos)(x)\, .
\ee
\item[(iv)] At this point only $d$ dimensional constant
curvature operators as well as $\partial_u$ (which
can be expressed in terms of the mass by 
Scherk--Schwarz reduction of the log-radial momentum) are left, so the theory 
is now
a $d$ dimensional one. However it still remains
to make the field redefinition~\eqn{redef} explained in
Section~\ref{gaugeanaly}, a computation easily performed
by first expressing $(\DIV,\GRAD)$ in the Casimir basis
$(\wt\DIV,\wt\GRAD)$ via~\eqn{undo}. The result, $S(\varphi_0,
\varphi_1,\wt\varphi_2,\wt\varphi_3)$ is invariant under 
the St\"uckelberg gauge transformations~\eqn{GG1},~\eqn{GG2} 
and~\eqn{stuck}. It may be viewed as a gauge invariant, St\"uckelberg, formulation
of the massive higher spin theory.
\item[(v)] The final step is to truncate to 
$S(\varphi_0,\chi)$ as described in the previous Section
(by either gauging away St\"uckelberg fields or decoupling
fields that do not transform under partial gauge variations)
and at this final juncture 
Scherk--Schwarz reduce  the log-radial momentum according 
to\footnote{Hermiticity of the action, in which terms 
linear in $\mu$ appear, follows
when the adjoint operation
$\dagger$ acts on the parameter $\mu$ as it does for the
derivative:
$\partial_u^\dagger=-\partial_u$. If one prefers 
an action in which no imaginary
parameters appear, the auxiliary field $\chi$ can be rescaled by a factor $i$,
at the cost of a ghostlike sign for its kinetic term. This is legal
because the auxiliaries are not part of the physical spectrum. Of course,
the main difficulty of constructing higher spin interactions is ensuring
that ghostlike auxiliaries are never physical excitations.} 
$\partial_u^2\equiv\mu^2$. At a fixed spin~$s$ for 
which $(\N-s)\varphi=0=(\N-s+3)\chi$,
the parameter $\mu$ is
\be\mu^2=-m^2+\Big(s+\frac{n}{2}-2\Big)^2\, ,\label{usualm}\ee
in terms of the more customary physical mass parameter $m$
(defined by requiring $m=0$ for strict masslessness).
\end{enumerate}
The above computations are straightforward yet tedious. Their result is
\bea
S&=&\varphi\ \Big[\ {\ss\square \ +\ \mu^2\ - \ (\Nt+\frac12)^2
\ +\ 2\ (\Nt-\frac n2+\frac12)(\Nt+\frac n2-\frac12)
\ -\ \frac14(\wt\GRAD-\G\wt \DIV)
\frac{1}{\Nt^2}\wt\DIV\ \Big] }
\ {\textstyle \delta_\TR}\ \varphi\nn\\[2mm]
&+&\varphi\ \Big[\ \ss{
\frac{4\ \Tt(\Tt-\frac12)}{(\Nt+\Tt)(\Tt-1)}\ 
\left(\square+2\ [\Tt-\frac n2+\frac12]
[\Tt+\frac n2-\frac12]
-\frac14\wt\GRAD \frac{\Tt-1}{\Tt(\Tt-\frac32)(\Nt-\Tt+2)}\wt\DIV
\right)}
\nn\\[1mm]&&\qquad\qquad\qquad\qquad\quad
{\ss \ - \ \left(\mu^2-[\Tt+\frac12]^2\right)
(\Nt-\Tt-1)} \Big]
\ {\ss 
\frac{\left(\frac{\Nt-\Tt+1}2\right)_{\!\sss \frac{\Nt+\Tt-2}2} }
{\left(\frac12\right)_{\!\sss \frac{\Nt+\Tt-2}2 }}}\ {\textstyle\Pi_{\TR}}
\  \varphi\nn\\[2mm]
&+&\!\!\! \ \Big\{2\ \varphi {\ss \G} \Big[ - 
{\ss\wt\GRAD  \ \frac{\mu-\Tt
+\frac32}{(\Nt+\Tt+2)}
\ +\  \G \
\frac{(\mu+\Tt-\frac32)(\Nt-\Tt+5)}{(\Nt-\Tt+2)(\Nt+\Tt+1)(\Nt+\Tt+4)}\wt\DIV\Big]}
\nn\\[1mm]&&
+\ \ \chi \, \ \Big[\ 
{\ss
\frac{4\ \Tt(\Tt-\frac12)}{\Tt-1}\left(\square+2
[\Tt-\frac n2+\frac12][\Tt+\frac n2-\frac12]
-\frac14\wt\GRAD
\frac{\Tt-1}{\Tt(\Tt-\frac32)(\Nt-\Tt+2)}
\wt\DIV
\right)}
\nn\\[1mm]&&\qquad
{\ss
\ -\  \left(
\mu^2-[\Tt+\frac12]^2\right)
(\Nt-\Tt+2)(\Nt+\Tt+3)}
\ \Big]
{\ss 
\frac{\Nt-\Tt+2}{\Nt+\Tt+1} } \Big\}
{\ss
\frac{\left(\frac{\Nt-\Tt+5}2\right)_{\!\sss\frac{\Nt+\Tt}2}}
{\left(\frac12\right)_{\!\sss\frac{\Nt+\Tt}2}} }
\ \chi\, .
\label{ACTION}
\eea
This is the generating function for all massive actions. 
When $(\N-s)\varphi=0=(\N-s+3)\chi$ it describes spin~$s$ 
excitations. For generic masses, the physical spectrum is correct
 because there is
a set of constraints whose leading
terms are all $s$ possible powers of the divergence $\DIV^t \, 
{\cal G}+\cdots =0$
of the equation of 
motion 
${\cal G}$~\cite{Deser:2001us,Deser:2001pe,Deser:2001wx,Deser:2001xr,Deser:2003gw,Deser:2004ji}. 
The existence of these 
constraints is guaranteed because the action~\eqn{ACTION}
arises from gauge fixing the St\"uckelberg
variations~\eqn{GG1},~\eqn{GG2}
and~\eqn{stuck}. At tuned values of the mass parameter,
the constraints become the Bianchi identities of partially
massless theories. Explicit formul\ae\  for these follow directly from
the partially massless gauge transformations presented in the 
next Section. Solving these constraints, leads finally to an
on-shell description where $\chi=0$ and, at spin~$s$,
\bea
\Big(\square-m^2+2(s-1)(s+n-2)\Big)\, \varphi=0=\DIV\varphi=\TR\varphi\, ,
\label{cshell}
\eea
which is the usual description of a massive higher spin field.
See~\cite{Deser:2003gw} for a representation theoretic account.

\section{Partially Massless Theories}

\label{part}

The simplest formulation of partially massless fields is 
on-shell  where one searches for residual gauge invariances
of the massive field equations~\eqn{cshell}. These exist at tuned
values of the mass as given in~\eqn{tunes}. This approach is described
in Appendix~\ref{onshell} and is actually a new result.

For interactions, one is more interested in an off-shell formulation
in terms of an action and accompanying gauge invariances, which we now 
describe. Essentially, all we need to do is reassemble
equations~\eqn{93}-\eqn{zeromode} for degeneracies of the massive
St\"uckelberg gauge transformations. Firstly, by Scherk--Schwarz
reduction, we focus our attention on the sector of fixed
log-radial momentum $\partial_u=\mu$. Then, a depth~$t$ degeneracy 
with traceless gauge parameter $\overline\xi_t$ appears
at the tuning
\be
\Big(\mu+\Tt+\frac12\Big)\overline\xi_t=0\, ,
\label{looney}
\ee
for gauge parameters
\be
\xi_1=\sum_{k=0}^{[\frac{t-1}{2}]} 
\G^{k}\ \Big[\prod_{j=2k+1}^{t-1}\Big(\frac{\overline u_j^{-1}v_j}{j-t}\ 
\Big)\Big]\overline\xi_t\, ,\quad
\xi_2=\sum_{k=0}^{[\frac{t}{2}]-1} \G^{k}\ \Big[\prod_{j=2k+2}^{t-1}
\Big(\frac{\overline u_j^{-1}v_j}{j-t}\Big)\Big]\overline\xi_t\, .
\label{params}
\ee
In these formul\ae\ $u_j=(j-t)\, \overline u_j$ and
\bea
\overline u_j&=&\left\{
\begin{array}{ll}
\qquad1&j=1\\[3mm]
\qquad2 &j=2,4,6,\ldots\\[3mm]
2j\, (\Tt-\frac j2+\frac12)\qquad&j=3,5,7,\ldots \, ,
\end{array}
\right.\\[4mm]
v_j&=&\left\{
\begin{array}{ll}
\ - \ \frac12 \ \wt\GRAD \, \frac1\Tt&j=1\\[3mm]
-2 \ \wt\GRAD\,  \frac{\Tt-1}{\Tt}\qquad&j=2,3,4,\ldots\, ,
\end{array}
\right. \
\eea
where the products are ordered from left to right with increasing $j$.
Then, with $\xi_1$, $\xi_2$ as stated, the partial gauge transformations
read
\bea
\delta\varphi&=&\GRAD\ \xi_1+\G\ \xi_2\nn\\[3mm]
&=&-\frac{1}{2}\ \wt\GRAD\ \frac{1}{\Tt}\, \xi_1
+\G\Big(\xi_2-\frac12\frac{1}{\Tt(\Nt-\Tt+2)}\wt\DIV\xi_1\Big)
\, ,\nn\\[2mm]
\delta\chi&=&-\TR\xi_1+\frac{1}{\TR\G}\TR\Big(
\GRAD\xi_2+(\mu-\Nt+\frac12)\ \xi_1
\Big)\nn \\
&=&
-\Big(1-\frac{\mu-\Nt-\frac32}{(\Nt+2)^2-\Tt^2}\Big)\TR\xi_1\nn\\
&-&\frac{1}{2}\, \Big( \frac{1}{(\Nt+2)^2-\Tt^2}\
\TR\, \wt\GRAD\ \frac{1}{\Tt}\
+\frac{1}{\Tt(\Nt-\Tt+2)}
\wt\DIV\Big)\, \xi_2\, .\nn\\
\eea

\subsection*{Examples}

\subsubsection*{Depth $t=1$ -- Strictly Massless Theory}
When $t=1$, we have $\xi_2=0$ and $\xi_1=\overline\xi_1$ is itself traceless.
Hence $\chi$ does not vary and 
\be
\delta\varphi=\GRAD\overline\xi_1\, ,
\ee
is the usual massless gauge transformation. 
It is a gauge invariance of the action~\eqn{ACTION}
when $\mu$ obeys the tuning condition~\eqn{looney}.
In terms of the usual physical mass parameter $m$ 
given in~\eqn{usualm} this means $m=0$. For generic
masses $m$ the action, being massive, is not gauge invariant
but the field equations are 
subject to the corresponding single derivative constraint
(see~\cite{Deser:2001us,Deser:2001pe} for a detailed explanation
of how gauge Bianchi identities become constraints in massive models).

Observe that the only fields 
transforming under this gauge variation are the doubly tracefree components
of $\varphi$, {\it i.e.}, $\varphi$ obeying $\TR^2\varphi=0$. This means
that we may set all other fields to zero. (In fact, since the decoupled fields
are in danger of being ghostlike, unitarity forces us to do so.)
Hence we obtain the usual description of a strictly massless higher
spin gauge field in terms of a doubly traceless field and a gradient
gauge transformation.

\subsubsection*{Depth $t=2$}
This is the first example of a higher (quadratic) derivative
gauge invariance. The tracefree part of $\xi_2=\overline\xi_2$
is the independent parameter and $\xi_1$ is determined by~\eqn{params}
to be
\be
\xi_1=-\frac{1}{2}\, \wt\GRAD \, \frac{1}{\Tt}\, \overline\xi_2\, .
\ee
The fields transform as
\bea
\delta\varphi&=&\GRAD\xi_1+\G\overline\xi_2\nn\\[2mm]
&=&
\frac{1}{4}\, 
\wt\GRAD\Big(\wt\GRAD+\frac12\G \wt\DIV\frac1{\Nt-1}\Big)\frac{1}{\Nt(\Nt+1)}\,  \overline\xi_2\nn\\
&+&\frac12\, \G\, 
\Big(\frac{\square-\frac12(n+1)(n-3)}{\Nt+1}+2\Nt\Big)\, \overline \xi_2
\nn\\[4mm]
\delta\chi&=&\frac14\, \frac1{\Nt(\Nt+1)}\, \wt \DIV \, \overline\xi_2\, .
\eea
For spin~2, the field $\chi$ is absent and we recover
the famous partially massless transformation $\delta\varphi=\GRAD^2\overline
\xi_2+\G\overline\xi_2$, or in pedestrian notation~\cite{Deser:1983tm,Deser:1983mm}
\be
\delta \varphi_{\mu\nu}=\Big(D_{(\mu} D_{\nu)}+g_{\mu\nu}\Big)\xi\,.
\ee
Again, this is a gauge invariance of the action~\eqn{ACTION}
when $\mu$ obeys~\eqn{looney} and otherwise
implies a double derivative constraint.
In terms of the usual, spin~$s$, physical, mass parameter the tuning is
at $m^2=2s+n-5$.

Finally, in accordance with the degree of freedom
counting arguments of~\cite{Deser:2001us}, we may truncate
most of the auxiliary fields because only the trace-free and doubly
trace-free parts of $\chi$ and $\varphi$ transform.

\section{Fermions}

\label{app1}

We define the space of symmetric spinors $\odot {\mathcal F}$
much as we did symmetric tensors. Elements of this space are spinor-tensor
fields of the form:
\be
\odot^s {\mathcal F} \ni \Psi=\psi_{\mu_1\ldots\mu_s}\ dx^{\mu_1}\ldots dx^{\mu_s}\, .
\ee
Again we make no restriction to elements of definite index content, so sums
of elements with differing values of $s$ are permitted.
All the operators that act  
on symmetric tensors, also act on symmetric spinors
with minor differences. In particular,
\be
\d_\mu:{\cal F}\rightarrow {\cal F}
\ee
operates in the same way as it does on symmetric tensors. It doesn't act on
the spinor tensor $\psi_{\mu_1\ldots\mu_s}$ and operates
on the differentials as in~\eqn{del}. The covariant derivative acts as per
equation~\eqn{covder}.
We use the Dirac matrices to construct several useful operators.
\begin{itemize}
\item{\it Gamma:}\be\GAMMA \equiv \g_{\mu} dx^{\mu} \, .\ee
Explicitly, \be \GAMMA \psi_{\mu_1\ldots\mu_s}dx^{\mu_1}\ldots dx^{\mu_{s}} 
= \GAMMA_{\mu_1} \psi_{\mu_2\ldots\mu_{s+1}}
    dx^{\mu_1}\ldots dx^{\mu_{s+1}} \, .\ee

\item{\it Gamma-trace:}\be \GAMMAC \equiv \g^{\mu} \d_{\mu} \, .\ee

This is the operator adjoint to $\GAMMA$. Explicitly,
\be \GAMMAC \psi_{\mu_1\ldots\mu_s}dx^{\mu_1}\ldots dx^{\mu_s} 
= s \GAMMA^{\mu} \psi_{\mu \mu_2\ldots\mu_s}dx^{\mu_2}\ldots dx^{\mu_s} \, .\ee
The operator $\N$ still counts the number or tensor indices, so its kernel, spinor fields,
coincides with that of~$\GAMMAC$.

\item{\it Dirac operator:}\be \DSLASH \equiv \g^{\mu}D_\mu \, . \ee
This is the usual Dirac operator on spinors and spinor-tensor fields.
\end{itemize}
Due to the Dirac matrix term in the 
covariant derivative commutator~\eqn{covder}
for spinors, it is
convenient to modify the $\DIV$ and $\GRAD$ operators. We define
\be\SDIV = \DIV + \frac{i}{2}\GAMMAC\ee
and
\be\SGRAD = \GRAD + \frac{i}{2}\GAMMA\, ,\ee
which results in the following commutation relation in terms of the old
commutators for $\DIV$ and $\GRAD$:
\be\left [ \SDIV, \SGRAD \right ] = \left [ \DIV, \GRAD \right ] + i\DSLASH \, .\ee
This definition is motivated by cosmological 
supergravity~\cite{Townsend:1977qa}, where the 
underlying anti de Sitter superalgebra implies the appearance
of a modified covariant derivative
\be
{\cal D}_\mu = D_\mu+\frac12\  \sqrt{-\Lambda/n}\  \gamma_\mu\, .
\ee
Since we work in de Sitter units $\Lambda=n$, explicit $i's $ appear
in formul\ae. Indeed, actions for de Sitter higher spin fermi fields
lose hermiticity, even though the underlying physical excitations
correspond to unitary, locally positive energy, representations 
of the de Sitter isometry group~\cite{Deser:2003gw}.
All our results can be easily translated to anti de Sitter space
by reinstating
the cosmological constant through dimensional analysis, taking care of
imaginary units as explained above.

\begin{figure}
\begin{center}
\epsfig{file=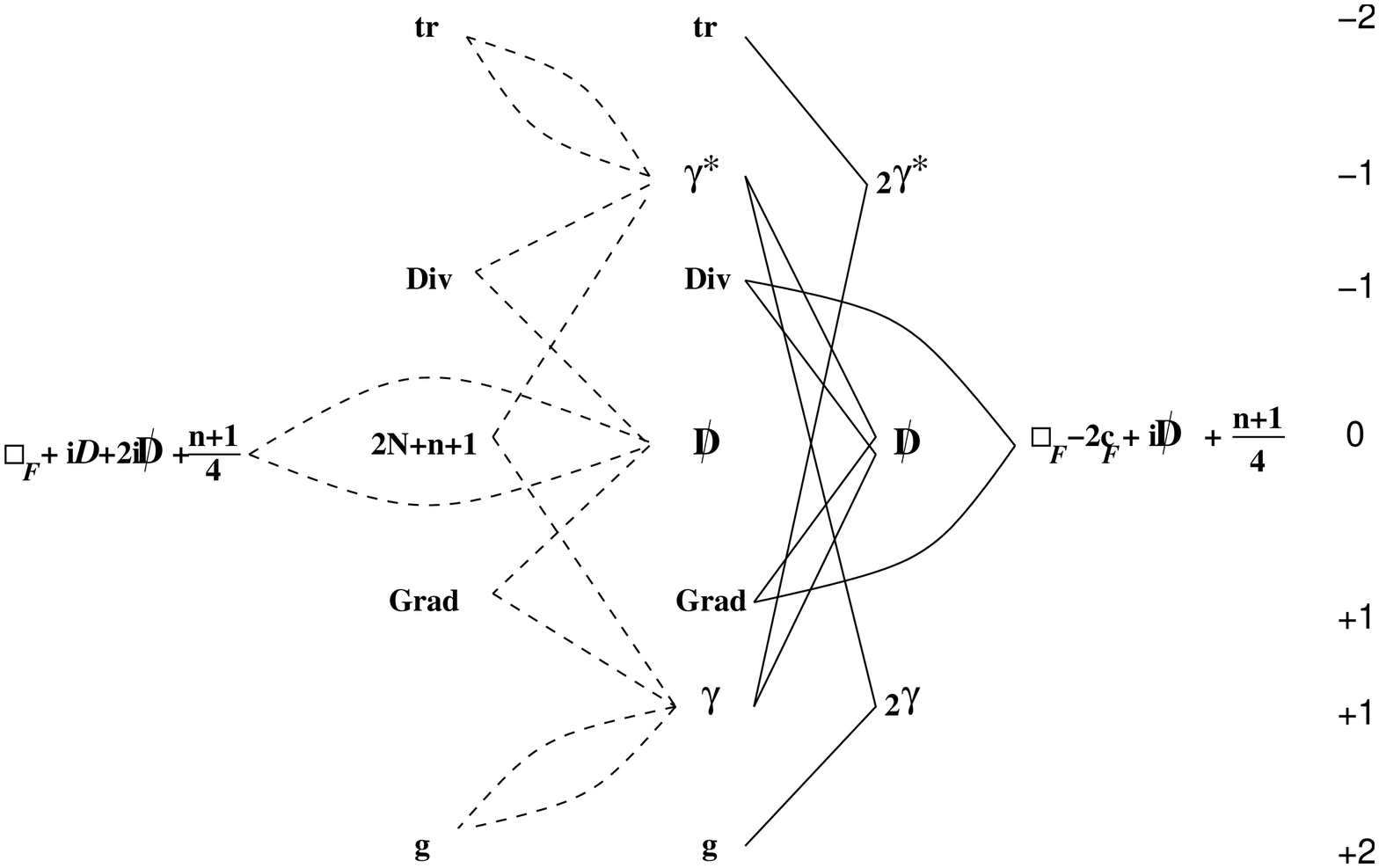,height=8cm}
\end{center}
\caption{The algebra of operators on
symmetric spinor tensors living on
$n+1$ dimensional constant curvature manifolds. The right-hand column
enumerates the weight with respect to the index operator $\N$.
Only non-vanishing or results different from the bosonic case in 
Figure~\ref{commutators} are shown. Dashed lines denote anticommutators,
while dashed loops are squares of operators (the diagram should be read from 
the middle column outwards). 
\label{supercommutators}}
\end{figure}

The commutation and anticommutation relations of the above 
operators with themselves and the bosonic operators mentioned
earlier are summarized in 
Figure~\ref{supercommutators}.
All the old bosonic commutation relations carry over except the
commutator of $\SDIV$ and $\SGRAD$.

Before we
describe the commutator of $\SDIV$ and $\SGRAD$ in more detail, let us
introduce two Casimirs and related operators
that appear in this superalgebra.
\begin{enumerate}
\item Denote the old $sl(2,\Real)$ Casimir $\c$ 
that commuted with $\TR$, $\G$, and $\N$ by
$\c_B$ (``$B$'' for bosonic). The corresponding operator for fermions, 
\be
\c_F = \c_B - \frac{1}{2}\left [ \GAMMAC, \GAMMA \right ]
  + \frac{1}{2}\left ( n+1 \right )\, ,\ee
commutes with $\TR$, $\G$, $\N$, $\GAMMA$, and $\GAMMAC$. (These
operators form an $so(2,1|1)$ super Lie algebra.)

\item The quadratic Casimir is the fermionic version of the bosonic Lichnerowizc
operator (denoted $\Box_B$ and
$\Box_F$, respectively):
\be \Box_F = \Box_B - \frac{1}{2}\left [ \GAMMAC, \GAMMA \right ]
   + \frac{1}{2}\left ( n+1 \right )\, .\ee
As in the bosonic case, the flat space limit is the Laplacian $\Delta$.
The operator $\Box_F$ is central.

\item There is a further quartic Casimir $\D$ of this
superalgebra first introduced in~\cite{Deser:2001us}, 
which we shall call the ``Lichnerowizc--Dirac operator'', 
given by
\be \D = (2\N+n-1)\DSLASH +2i\c_F 
  -2(\SGRAD \GAMMAC + \GAMMA \SDIV) + 2 \GAMMA \DSLASH \GAMMAC\, .\ee
The Lichnerowizc--Dirac operator is also central -- it {\it commutes}
with every other generator.
\end{enumerate}
With the above technology on board, a simple expression for the 
$\SDIV$, $\SGRAD$ commutator, based on~\eqn{covder}, follows
\be
[\SDIV,\SGRAD]=\Box_F-2\c_F+i\DSLASH+\frac{n+1}{4}\, .
\ee 
In turn, the square of the Dirac operator reads
\be
\frac12\{\DSLASH,\DSLASH\}=\Box_F+i\D +2i\DSLASH+\frac{n+1}{4}\, .
\ee
The appearance of these various Casimirs is simply explained
by employing a two component notation in which the $sl(2,\Real)$
subalgebra action is made manifest. See Appendix~\ref{twoC}
for details. Also, a Casimir basis reformulation of the 
fermionic constant curvature algebra ought to exist, but we reserve 
its development for future study.

\subsection{Generating Function for Massless Fermion~Actions}
\label{fermions}

We now give the half integer spin generalization of the massless,
constant curvature
action generating function. The minimum covariant field content for massless,
half integer, higher spins is a traceless-gamma-traceless symmetric 
spinor-tensor $\psi$:
\bea
\GAMMAC \TR \psi = 0.
\eea
The correct degrees of freedom are imposed through the gauge invariance
\bea
\delta \psi = \SGRAD \xi\, .\label{m=Fgauge}
\eea
Here, the parameter $\xi$ is gamma-traceless
\bea
\GAMMAC \xi = 0\, .
\eea
The inner product is defined much as for the bosonic case
\bea
\langle \xi_1, \xi_2 \rangle = \int \sqrt{-g}\ \overline\xi_1 \xi_2
\eea
where the adjoint operation defined earlier for tensor operators is
extended in the natural way to the fermionic case.

The action for massless fields (assuming single derivatives) is then uniquely
determined by the gauge invariance~\eqn{m=Fgauge} which yields
\bea
S &=& \int \overline\psi \ 
\Big\{\D +i(\N^2+n\N-4) - i\c_F  
       + \G \SDIV \GAMMAC+ \GAMMA \SGRAD \TR\nn
      \\
   &&  - \frac{1}{2} \G \DSLASH \TR  
       -i \GAMMA \Big(4\N +2n-3\Big)\GAMMAC
  - \DSLASH (2\N+n-3)
       \Big\} \ \psi
  \ \equiv  \  \int \overline\psi \cal{R} \psi.\nn\\
\eea
It is easy to check the Bianchi identity $\SDIV {\cal R}=0$ 
mod $\GAMMA$
corresponding to the gauge invariance~\eqn{m=Fgauge}.

\section{Scalar Fields on the Total Space}

\label{total}

The ideas in this Section apply to massive or massless
theories in both flat and constant curvature spaces. 
However, for simplicity of presentation, we
concentrate on the case of massless higher spins in $d$ dimensional
Minkowski space $M$. The generating function for their actions is
\be
S=\int\Phi\, \{{\ss
\square-\GGRAD\DDIV+\frac12[\GGRAD^2\TTR+\GG\DDIV^2]
-\frac12\GG[\square+\frac12\GGRAD\DDIV]\TTR
}\}\, \Phi\,  .
\ee
Here the field $\Phi$ is subject to the double traceless
constraint $\TTR^2\Phi=0$. Consider now the total space $E$
of the cotangent bundle $T^*M$ with coordinates $(x^\mu,dx^\mu)$.
In the above action the field $\Phi=\Phi(x,dx)$ and may therefore
be viewed as a scalar field on $E$. Hence, the 
theory of massless higher spins is reduced to a scalar theory
in twice as many dimensions.

If we call $y^\mu\equiv dx^\mu$ our action becomes
$$
\!\!\!\!\!\!
S=\int_E \Phi({\ss x,\frac{\partial}{\partial y}})\ 
\{{\ss
\square_x 
- y^\mu \frac{\partial}{\partial x^\mu}
  \frac{\partial}{\partial x_\nu} \frac{\partial}{\partial y^\nu}
+\frac12([y\cdot \frac{\partial}{\partial x}]^2 
 \square_y
 +y\cdot y[\frac{\partial}{\partial x}\cdot \frac{\partial}{\partial y}]^2 
 )
}
$$
\be
\qquad\qquad\qquad\qquad
{\ss
-\frac12 y\cdot y (\square_x+\frac12
  y^\mu \frac{\partial}{\partial x^\mu}
  \frac{\partial}{\partial x_\nu} \frac{\partial}{\partial y^\nu}
 )\square_y
}
\} \ 
\Phi({\ss x,y})
\ee
where the integral over the total space $E$ is integration
over the base space $M$ and setting $y$ to zero (after
performing all possible $y$ derivatives).

Clearly this is a rather peculiar scalar field theory,
since it has (i) a non-standard integration measure,
(ii) a constraint $\square_y^2\Phi=0$ and (iii) a gauge invariance
$\delta\phi=y^\mu \frac{\partial}{\partial x^\mu} \xi$.
Moreover it is higher derivative and position dependent.
These apparent drawbacks are a necessary price for handling all 
spins simultaneously. Possibly, a clever choice of gauge 
might lead to a tractable analysis, but we leave this issue to further study. 

\section{Conclusions}

\label{conc}

This Article dealt with two main achievements. The first was
a detailed analysis of the algebra of operators acting
on the symmetric tensor bundle $\odot T^*M$ for an arbitrary
manifold $M$. For general manifolds $M$, this algebra is
a deformation of the semi-direct product
$sl(2,\Real)\, \raisebox{.6mm}{$\sss |$}
\hspace{-2.15mm}\times\Real^2$ Lie algebra where the
deformation appears in commutators
of the $\Real^2$ generators. In particular, by enlarging the
universal enveloping algebra to include the square root
of the $sl(2,\Real)$ Casimir and rational functions thereof,
we displayed an operator-valued diagonalization of
the $sl(2,\Real)$ action on the
$\Real^2$ factor. This result holds already in the undeformed
Lie algebra and provides an extremely useful calculus for
universal enveloping algebra valued computations.

In the case that the underlying manifold $M$ is constant curvature,
the deformation is extremely simple, given by the sum of the central
Lichnerowicz wave operator and the quadratic $sl(2,\Real)$ Casimir.
In some sense, we may view these results as the totally symmetric
generalization of differential forms. In particular it 
should provide a powerful framework for future studies of
constant curvature manifolds.

An obvious generalization of these results is to the tensor bundle
${\cal T}M$ of all possible tensors over the manifold $M$. 
These can be analyzed in terms of Young diagrams where each 
row corresponds to a symmetric tensor representation and in
turn a copy of the algebra discussed above. This algebra must
then be enlarged by operators mixing rows. We expect a similar
picture to emerge, but leave this avenue to future studies.

A central physics application of totally symmetric
tensor fields is the theory of higher spins--the second topic of
this Article. Here the advances are as follows 
\begin{itemize}
\item Computations involving
symmetric tensors on curved manifolds are vastly simplified
which is an  important technical advance.
\item
Working in the space of all symmetric tensors
we can write generating functions for all higher spin theories,
rather than working at any given spin. 
\item 
We have shown how the St\"uckelberg formalism of~\cite{Zinoviev:2001dt} 
arises naturally via radial dimensional reduction. In particular,
our formalism is in terms of a minimal set of unconstrained
fields, rather than cumbersome towers of auxiliary fields.
\item
Partially massless gauge transformations can be simply understood
as degeneracies of the flat space transformation
$\delta\Phi=\GGRAD\Xi$. This allows us to write 
explicit formul\ae\  for partially massless gauge transformations
at arbitrary spin.
\end{itemize}

We end with some more speculative remarks. Firstly, since our
formalism deals naturally with arbitrary spins including infinite
towers of higher spins, it seems that it should
be well suited to studying the higher spin interaction problem.
Indeed there are already indications~\cite{Vasiliev:2000rn,Brink:2000ag} 
that infinite
towers of spins are a necessity for consistency of interactions. Moreover
higher massive String states provide consistent couplings of
infinite towers of higher spins. Clearly it would be very desirable
to formulate interacting String dynamics within the approach
we have presented, perhaps generalized to allow for tensors of mixed
symmetry type, as discussed above.

Our final comment concerns the total space of the symmetric tensor
bundle $\odot T^*M$. As remarked earlier we may view sections of this
bundle as functions of coordinate differentials. Hence, as discussed  
in Section~\ref{total}, higher spins can be described in terms
of a novel, total space, scalar field theory. Clearly this approach
deserves further investigation and also cries out for a String Field
theoretic formulation.

\section*{Acknowledgments}
It is a pleasure to thank Stanley Deser, Rod Gover, Andrew Hodge, Misha Khovanov, Sven Moch,
Misha Movshev and Albert Schwarz for pertinent discussions.
This work was supported by the 
National Science Foundation under grant PHY01-40365.

\begin{appendix}

\section{The Constant Curvature Algebra}
\label{harm}

\subsection{Non-Commutative Harmonic Oscillators}
\label{ncho}

There is an interesting relationship between the constant
curvature algebra
and the harmonic oscillator. Begin with the flat space algebra 
and make the following identifications
\bea
\TR \longleftrightarrow-\frac1\hbar p^2&&\nn\\
&\DIV \longleftrightarrow -ip&\nn\\
\N \longleftrightarrow \frac{i}{2\hbar}\ (xp+px)
&&\square \longleftrightarrow \hbar\nn\\
&\GRAD \longleftrightarrow -x&\nn\\
\G \longleftrightarrow \frac1\hbar x^2\, .
\eea
Then, with the usual quantum commutator $[p,x]=-i$,
the two algebras coincide. The right hand side is the
spectrum generating $sl(2,\Real)$ algebra of the harmonic oscillator
with hamiltonian $2 H=\hbar(\G-\TR)$ corresponding to the generator
of the $SO(2)$ maximal compact subgroup.

This observation was not employed in our higher spin
investigations because
the harmonic oscillator representation diagonalizes the Casimir
$\c=\frac12 n(n-2)$, in particular there is no obvious constant curvature
deformation. Nonetheless, it is conceivable that, using the reverse logic,
the constant curvature algebra may provide an interesting non-commutative
deformation of the harmonic oscillator algebra.

\subsection{Higher Dimensional Embedding}
\label{in_bed}

The constant curvature algebra in dimension $d=n+1$, can be
embedded in the flat space algebra of one dimension higher.
This is achieved as follows: Introduce new operators $du$,
$\partial$, $\partial_u$ and $e^u$ with non-vanishing commutators
\be
[\partial,du]=1\, ,\qquad [\partial_u,e^u]=e^u\, .
\ee
Then, the flat space algebra
$\{\TTR,\DDIV,\square_{d+1},\N_{d+1},\GGRAD,\GG \}$ in $d+1$ dimensions 
can be written in terms of its $d$ dimensional constant curvature 
counterpart as
\bea
{}\TTR=e^{-2u}(\partial^2+\TR)&&\nn\\[2mm]
&\hspace{-1.8cm}
\DDIV=e^{-2u}(\DIV-\TR\ du+\partial\ [\partial_u+n+1-du\partial])&\nn\\[2mm]
\N_{d+1}=du\partial +\N
&&
\hspace{-5.0cm}\square_{d+1}=[\DDIV,\GGRAD]\, \quad\mbox{(central)}\nn\\[2mm]
&\hspace{-2.8cm}\GGRAD=\GRAD+\G\ \partial+du\ [\partial_u-2\N-du\partial]&
\nn\\[2mm]
\GG=e^{2u}(du^2+\G)\, .&&
\label{algembed}
\eea
Here
$$
\square_{d+1}\; =\;
e^{-2u}\Big(\square+\partial_u^2+[n-2N-2du\partial]\partial_u
+2[\GRAD\partial-\DIV du]$$
\be
+\G\partial^2+\TR du^2 
+du(du\partial-2n)\partial-\c+\N(\N-n-1)\Big)\, ,
\ee
so it might seem remarkable that these
complicated operators obey the simple flat space
algebra -- a point that is clear based, either on
explicit computations or our discussion of
radial dimensional reduction in Section~\ref{radial}.

We also note that there is the obvious relation between flat space
algebras in adjacent dimensions
\bea
{}\TTR=\partial^2+\TR&&\nn\\
&\DDIV=\DIV+\partial\ \partial_u&\nn\\[2mm]
\N_{d+1}=du\partial +\N&&\square_{d+1}=\square+\partial_u^2\nn\\
&\GGRAD=\GRAD+du\ \partial_u&\nn\\
\GG=du^2+\G\, .&&
\eea

\subsection{Two Component Notation}
\label{twoC}

We can also define a two component notation for 
the constant curvature algebra that 
makes the $sl(2,\Real)$ subalgebra manifest.
This notation also generalizes to the fermionic
superalgebra of Section~\ref{app1}, so we concentrate on that
case.

Let Greek indices
$\alpha,\beta,\ldots=1,2$ label the
fundamental representation of $sl(2,\Real)$ and define
\be
f^{\alpha \beta} = \left(
  \begin{array}{cc}  \G & \N+\frac{n+1}{2}\\
       \N+\frac{n+1}{2} & \TR \end{array} \right)=f^{\beta \alpha }\, ,
\qquad
\epsilon^{\alpha \beta} = \left(
  \begin{array}{cc}
    0 & 1\\
    -1 & 0\\
  \end{array} \right )=\epsilon_{\alpha\beta}\, ,
\ee
along with
\be
v^{\alpha} = \left(
  \begin{array}{c}
    \SGRAD\\
    \SDIV \\
  \end{array} \right )
\, ,\qquad
\GAMMA^{\alpha} = \left(
  \begin{array}{c}
    \GAMMA\\
    \GAMMAC \\
  \end{array} \right )\, .
\ee
Then $sl(2,\Real)$ acts on a single index of an arbitrary two component vector
as:
\bea
\begin{array} {lll}
\left [f^{\alpha \beta}, X^{\gamma} \right ] & = &
  \epsilon^{\gamma \alpha} X^{\beta} +
  \epsilon^{\gamma \beta} X^{\alpha}\nonumber \\ 
& = & 2 \epsilon^{\gamma (\alpha} X^{\beta)}\, , 
\end{array}
\eea
which implies
\bea
\left [f^{\alpha \beta}, f^{\gamma \delta} \right ] = 
  \epsilon^{\gamma \alpha} f^{\beta \delta} +
  \epsilon^{\gamma \beta} f^{\alpha \delta} +
  \epsilon^{\delta \alpha} f^{\gamma \beta} +
  \epsilon^{\delta \beta} f^{\gamma \alpha}\, .
\eea
We also have
\bea
\left \{\GAMMA^{\alpha}, \GAMMA^{\beta} \right \} = 
 f^{\alpha \beta}\,, 
\eea
as well as
\bea
\left [\GAMMA^{\alpha}, v^{\beta} \right ] = 
 -\epsilon^{\alpha \beta} \DSLASH 
&&
\left(\Rightarrow
\DSLASH=
-\frac12( \GAMMA^\alpha v^\beta- v^\beta\GAMMA^\alpha)\epsilon_{\alpha\beta}\right)\, ,\nn\\
\left \{\GAMMA^\alpha,\DSLASH \right \}&=&2v^\alpha\, .\nn\\
\eea
Using
$
[\SDIV, \SGRAD]=v^{\alpha}\epsilon_{\alpha \beta} v^{\beta}
$, we obtain explanatory formul\ae\ for the various Casimirs
\bea
\c_F&=&-\frac12 f^{\alpha\beta}\epsilon_{\beta\gamma}f^{\gamma\delta}
\epsilon_{\delta\alpha}
+\gamma^\alpha\epsilon_{\alpha\beta}\gamma^\beta+
\frac14(n+1)(n-3)\, ,\\
\Box_F&=&-v^\alpha\epsilon_{\alpha\beta} v^\beta-i\DSLASH+2\c_F-\frac{n+1}{4}
\, ,\\
\D\ &=&\GAMMA^\alpha\DSLASH \epsilon_{\alpha\beta}\GAMMA^\beta+2i\c_F.
\eea
The higher spin actions we 
consider do not manifest this $sl(2,\Real)$ symmetry, so little
use was  made of this elegant notation.

\section{On-shell Partially Massless Fields}

\label{onshell}

There are various equivalent
descriptions of partially massless fields. The simplest
is the ``on-shell'' one. A spin~$s$, depth~$t$,
on-shell, partially massless field is a section $\varphi$ of $\odot T^*M$ 
subject to 
\begin{itemize}
\item[(i)] $s$-index symmetric tensor: $$\N\varphi=s\varphi$$
\item[(ii)] trace-free: $$\TR\varphi=0$$
\item[(iii)] divergenceless: $$\DIV\varphi=0$$
\item[(iv)] eigenmode of Laplacian\footnote{Or for Lichnerowicz devotees
$\Big[\square+2\N(\N+n-t-2)+(t+1)(t+1-n)\Big]\varphi=0$.}:  
     $$\Big[\Delta+(\N-t)(\N+n-t-2)-\N-n+1\Big]\varphi=0$$ 
\end{itemize}
A generic eigenvalue of the Laplacian
corresponds to a massive field. The above value is special
because the system then enjoys a residual gauge invariance.
For example, when $s=t=2$ we have
\be
[\square+n+1] h=0=\DIV h=\TR h\, .\label{os2}
\ee
If we make a variation $\delta h=(\GRAD^2+\G)\, \overline\xi_2$ we find
$$
[\square+n+1] \delta h=(\GRAD^2+\G)[\square+n+1]\, \overline\xi_2\, ,\quad
\DIV\delta h=\GRAD[\square+n+1]\, \overline\xi_2\, ,$$\be
\TR\delta h=2[\square+n+1]\, \overline\xi_2\, .
\ee
Namely, the on-shell equations~\eqn{os2} enjoy a residual gauge invariance
when $[\square+n+1]\, \overline\xi_2=0$. This equivalence relation of
on-shell fields $h$ in turn implies the correct physical degree of
freedom count $d(d+1)/2-d-1-1=d(d-1)/2-2$.

The same computation can be performed in general.
On-shell, 
the depth~$t$, residual, partially massless 
gauge transformations are most simply given as
\be
\delta\varphi=\delta_{\TR}\  \wt {\GRAD}^{\ t} \ \, \overline\xi_t.
\ee
These transformations obviously obey the trace and
Laplace conditions (ii) and (iv) whenever the parameter $\, \overline\xi_t$ 
does.
Checking the divergence condition~(iii) requires a more involved computation
using the $\wt\DIV\, \wt\GRAD$ relation~\eqn{casalg}. That it holds precisely
is in fact an excellent cross check of that algebra.

%
%

\end{appendix}

\end{document}